\documentclass[useAMS,usenatbib,twocolumn]{mn2e}
\usepackage{graphicx,amssymb}
\usepackage{subfigure,appendix,enumerate}
\usepackage{url,hyperref,xcolor}
\usepackage{multirow}
\usepackage{lscape}
\usepackage{natbib}

\def\apj{\ifmmode ApJ \else ApJ \fi}    
\def\apjl{\ifmmode  ApJ \else ApJ \fi}    %
\def\apjs{\ifmmode  ApJS \else ApJS \fi}
\def\aap{\ifmmode A\&A \else A\&A\fi}
\def\aaps{\ifmmode A\&AS \else A\&AS\fi}    %
\def\mnras{\ifmmode MNRAS \else MNRAS \fi}    %
\def\nat{\ifmmode Nature \else Nature \fi}
\def\prl{\ifmmode Phys. Rev. Lett. \else Phys. Rev. Lett.\fi}
\def\prd{\ifmmode Phys. Rev. D. \else Phys. Rev. D.\fi}
\def\pasp{\ifmmode  PASP \else PASP \fi}
\def\gca{\ifmmode Geochimica Cosmochimica Acta \else Geochimica Cosmochimica Acta \fi}
\def\araa{\ifmmode ARAA \fi}
\def\aapr{\ifmmode A\&ARv \fi}
\def\aj{\ifmmode AJ \else AJ \fi}

\definecolor{dark-green}{rgb}{0.1,0.49,0.4}
\hypersetup{colorlinks=true, urlcolor=blue, citecolor=dark-green}

\usepackage{color}

\begin{document}
\title[EXO 0748$-$676 in quiescence]{The cooling, mass and radius of the neutron star in EXO 0748$-$676 in quiescence with XMM-Newton} 

\author[Z. Cheng et al.]
{Zheng Cheng$^1$\thanks{Email: zheng@astro.rug.nl}, 
Mariano M\'{e}ndez$^1$, 
Mar\'{\i}a D\'{\i}az-Trigo$^2$,
Elisa Costantini$^3$,
\\ 
$^1$Kapteyn Astronomical Institute, University of Groningen, Postbus 800, 9700 AV Groningen, The Netherlands \\
$^2$ESO, Karl-Schwarzschild-Strasse 2, 85748 Garching bei M\"unchen, Germany\\
$^3$SRON, Netherlands Institute for Space Research, Sorbonnelaan 2, 3584 CA Utrecht, The Netherlands\\ 
}

\maketitle
\begin{abstract}

We analyse four XMM-Newton observations of the neutron-star low-mass X-ray binary EXO 0748$-$676 in quiescence. 
We fit the spectra with an absorbed neutron-star atmosphere model, without the need for a high-energy (power-law) component; 
with a 95 per cent confidence the power-law contributes less than 1 per cent to the total flux of the source in $0.5-10.0$ keV. 
The fits show significant residuals at around 0.5 keV which can be explained by either a hot gas component around the neutron star or a moderately broad emission line from a residual accretion disc. 
The temperature of the neutron-star has decreased significantly compared to the previous observation, from 124 eV to 105 eV, with the cooling curve being consistent with either an exponential decay plus a constant or a (broken) power law. 
The best-fitting neutron-star mass and radius can be better constrained if we extend the fits down to the lowest possible energy available. For an assumed distance of 7.1 kpc, the best-fitting neutron-star mass and radius are $2.00_{-0.24}^{+0.07}~M_\odot$ and $11.3_{-1.0}^{+1.3}$ km if we fit the spectrum over the $0.3-10$ keV range, but $1.50_{-1.0}^{+0.4}~M_\odot$ and $12.2_{-3.6}^{+0.8}$ km if we restrict the fits to the $0.5-10$ keV range.
We finally discuss the effect of the assumed distance to the source upon the best-fitting neutron-star mass and radius. As systematic uncertainties in the deduced mass and radius depending on the distance are much larger than the statistical errors, it would be disingenuous to take these results at face value.

\end{abstract}
 

\section{Introduction}
\label{intro}

EXO 0748$-$676 is an X-ray transient that was discovered with the European 
X-ray Observatory (EXOSAT) in 1985 \citep{parmar1986}. 
The source exhibits 8.3-minute duration X-ray eclipses that recur with a 
period of 3.82 hours. 
The periodic X-ray eclipses and irregular 
X-ray dips indicate that the system is a low-mass X-ray binary (LMXB) 
at a relative high inclination angle.
\cite{parmar1986} estimated that the inclination of the system lies in 
the range 75$^\circ - 82^\circ$. 
\cite{gottwald1986} detected type \uppercase\expandafter{\romannumeral1} 
X-ray bursts in this source, and therefore identified the compact object 
as a neutron star (NS). \cite{homan2000} reported a 695-Hz quasi-periodic 
oscillation (QPO) in an observation with the Rossi~X-ray~Timing~Explorer 
(RXTE). \cite{villarreal2004} detected a 45-Hz oscillation in the average 
power spectrum of 38 X-ray bursts, which they interpreted as the spin 
frequency of the NS. \cite{galloway2010} detected a 552-Hz oscillation 
in the rising phase of two type \uppercase\expandafter{\romannumeral1} 
X-ray bursts, and concluded that this is the spin frequency of EXO 0748$-$676, 
rather than the 45-Hz oscillation.

\cite{wolff2005} detected a few unusually strong X-ray bursts 
with a peak flux that was approximately 4 times higher than that 
of the normal bursts. They interpreted this as photospheric 
radius expansion (PRE) bursts, which indicated that the Eddington 
luminosity was reached at the peak of those bursts.
Assuming a typical NS mass of 1.4 
M$_{\odot}$, \cite{wolff2005} derived a distance to EXO 0748$-$676 of 7.7 
kpc for a helium-dominated photosphere, and 5.9 kpc for a hydrogen-dominated photosphere. By considering the touchdown flux and 
the high orbital inclination of the system, 
\cite{galloway2008a} estimated a distance of 7.1 $\pm$ 1.2 kpc 
(see also \citealt{galloway2008b}).

In 2008, observations with the Proportional Counter Array 
on-board RXTE \citep{wolff2008a} and the Swift~X-Ray~Telescope 
\citep{wolff2008b} showed that the 
X-ray flux of EXO 0748$-$676 was declining, indicating that 
accretion was ceasing and the source was transitioning to quiescence 
after accreting for over 24 years. This made it possible to study 
the cooling process of this source \citep{degenaar2011,guobao2011,diaz2011,degenaar2014}.

Isolated NS mainly cool via neutrino emission from the stellar core and 
photon emissions from the surface \citep{yakovlev2004}. 
For a quasi-persistent X-ray binary like EXO 0748$-$676, 
the outburst phase can last from years to decades, and pycnonuclear 
reactions can cause a significant temperature 
gradient between the core and the crust \citep{degenaar2011}.
When accretion ends, the crust is expected to thermally relax on a time 
scale of years, which can provide us with
information on the properties of the NS crust \citep{haensel2008,brown2009}.

In this paper we present the spectral analysis of four XMM-Newton observation 
of EXO 0748$-$676 in quiescence. The first three of these observations
were already analysed by \citet{diaz2011}, while the last observation 
was obtained about three years after the previous one, providing a long baseline to 
study the cooling process of EXO 0748$-$676. We describe 
the details of the observations and the data reduction and analysis 
in \S\ref{data}, we show the results in \S\ref{results}, and we discuss our 
findings in \S\ref{discussion}.

\section{Data reduction and analysis}\label{data}

\begin{table}
\caption{XMM-Newton observational data log of EXO~0748$-$676}
\label{table1}
\begin{tabular}{ c c c c}
\hline
\hline
Observation ID & Date   & Instrument & Exposure time (ks)\\
           \hline
0605560401 & 2009-03-18 & PN   & 36.27\\
           &            & MOS1 & 43.15\\
           &            & MOS2 & 42.04\\
           &            & RGS1 & 47.00\\
           &            & RGS2 & 45.44\\
           \hline
0605560501 & 2009-07-01 & PN   & 86.16\\
           &            & MOS1 & 99.60\\
           &            & MOS2 & 99.60\\
           &            & RGS1 & 101.20\\
           &            & RGS2 & 101.30\\
           \hline
0651690101 & 2010-06-17 & PN   & 24.80\\
           &            & MOS1 & 29.62\\
           &            & MOS2 & 29.98\\
           &            & RGS1 & 30.22\\
           &            & RGS2 & 30.36\\
           \hline
           &            & PN   & 58.88\\
           &            & MOS1 & 67.44\\
           &            & MOS2 & 67.48\\
           &            & RGS1 & 70.78\\
           &            & RGS2 & 70.74\\
           \hline        
0690330101 & 2013-04-15 & PN   & 91.32\\
           &            & MOS1 & 103.00\\
           &            & MOS2 & 103.10\\
           &            & RGS1 & 104.20\\
           &            & RGS2 & 104.30\\
\hline\\
\end{tabular}
\end{table}

EXO 0748$-$676 was observed with 
XMM-Newton on 2008 November 6 at 08:30$-$16:42 UTC, 
just after it turned into quiescence 
\citep[obsID 0560180701, see also][]{guobao2011}, 
and subsequently four more times from March 18 2009 to April 15 2013. 
\cite{guobao2011} found that, besides the thermal emission from the 
neutron star, in the first observation there is a significant contribution 
of a non thermal component. \cite{diaz2011} found that the non-thermal 
component was not present in the following three observations. Here we find 
(see below) that the same is true in the last observation and, therefore, 
in the rest of the paper we only analyse the last four XMM-Newton observations 
of this source.

Data were collected simultaneously with the European Photon Imaging Cameras 
\citep[EPIC;][]{struder2001} and the Reflection Grating Spectrometers \citep[RGS;][]{Denherder2001}.
The EPIC cameras
consist of one PN and two Metal Oxide Semi-conductor 
(MOS) detectors, which offer sensitive imaging observations over a 
field of view of 30\arcmin and an energy range from 0.15 to 12 keV, while the two RGS instruments
cover the range between 0.3 and 2.0 keV with a resolution of 100$-$500.
We give the log of the observations in Table \ref{table1}. 

We reduced the XMM-Newton Observation Data Files using the Science 
Analysis Software (SAS) version 14.0.0. We extracted the event files for 
the PN and the two MOS detectors using the tasks EPPROC and EMPROC, 
respectively, and we processed the RGS data using the task RGSPROC; following 
the recommendations of the XMM-Newton team, in all cases we used the default 
parameters of these tasks. Since 
the source was very weak during all these observations, the RGS instruments 
collected very few photons (both RGS combined contained less than 5\% of the 
photons of the PN and MOS instruments combined) and, therefore, in the rest of the 
paper we only fit the PN and MOS data. We, however, checked that the best-fitting 
parameters do not change significantly if we also include the RGS data to the fits. 
We mention this when necessary.

In order to identify the existence of possible flaring particle 
background, we extracted light curves for energies 
larger than 10 keV for MOS and in the 10$-$12 keV range for PN. We found 
soft proton flares in the MOS data of observation 0651690101 
and in small parts of the PN data of all observations. 
Following \cite{piconcelli2004}, we calculated the cumulative signal-to-noise ratio, SNR, as a function of the background count rate for all cameras in all observations. We found that in all cases the cumulative SNR increases monotonically up to the highest background count rates, indicating that, for these observations, it is better not to filter out the particle background flares. Nevertheless, to assess the effect of the background flares upon our analysis, we defined good intervals, GTI, as the times in which the count rate in the above bands was below 0.35 counts $\rm{s}^{-1}$
for MOS and 0.40 counts $\rm{s}^{-1}$ for PN. We applied these 
GTI files, together with standard filters, to the event data, and 
we subsequently removed the exposure time range during the eclipses (see \S\ref{intro}).
We also extracted spectra without removing the flaring time (but excluding the eclipses), 
and found that these spectra were consistent with those for which we had excluded 
the flares. 
Therefore, in the rest of the paper we do not filter out the flaring 
time in the data.

We extracted source spectra in the 0.3$-$10.0 keV band from a circular region 
with a radius of 35 arcsec centred on the source, and the 
corresponding background spectra from a source-free region with 
a radius of 35 arcsec, using the SAS task EVSELECT. We confirmed 
that the data were not piled up using the task EPATPLOT. We created 
the photon redistribution matrices (RMF) and ancillary response file (ARF), and 
re-binned the spectra to have at least 25 counts per channel using the task 
GRPPHA.

\begin{figure}
\begin{center}
\includegraphics[angle=270,width=8cm]{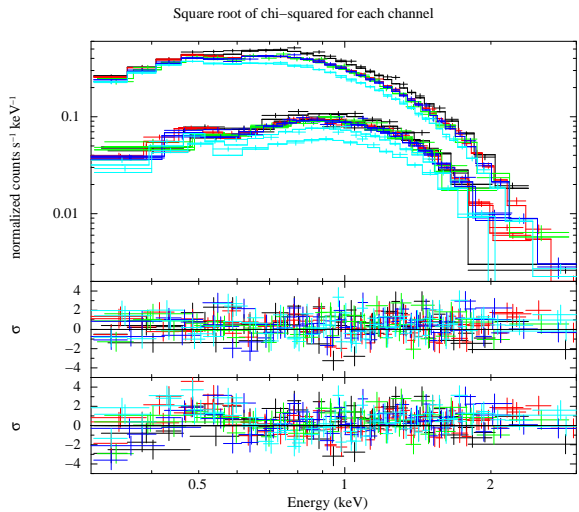}
\caption{
The 15 XMM-Newton EPIC spectra of EXO 0748$-$676 with the best-fitting
model. Top panel: Fits with the model {\sc const * phabs * (nsatmos + vapec)} in
the $0.3-10$ keV range. The best-fitting NS mass and radius are 
$M_{\rm ns}$=$2.05^{+0.09}_{-0.39}$ M$_{\odot}$ and $R_{\rm ns}$=$11.4 \pm 2.1$ 
km, respectively. 
The best-fitting Nitrogen bundance $Z_{\rm N}=21\pm8$, the temperature of the hot gas for the 4 observations are 269$\pm$43 eV, 172$\pm13$ eV, 250$\pm$29 eV and 152$\pm$16 eV, respectively. 
(All errors represent the 90\% confidence interval of the given parameter for a single interesting parameter.) For plotting 
purposes we have rebinned the data by a factor 8. 
Middle panel: The residuals, (data$-$model)/error, with respect to the best-fitting model.
Bottom panel: The fitting residuals of the best-fitting model after removing the {\sc vapec} component. 
We plot the PN, MOS1 and MOS2 spectra and models of each observation 
listed in Table \ref{table1} in black, red, green, blue and light blue, 
respectively.
}
\label{spectra1}
\end{center}
\end{figure}

We fitted the spectra in the 0.3$-$10.0 keV band using XSPEC 
version of 12.8.2. We fitted all four observations, in total 15 
spectra, simultaneously (see Table \ref{table1}).
We modelled the spectra using the NS hydrogen atmosphere 
model {\sc nsatmos} \citep{heinke2006}. This model covers a wide range of surface gravity and 
effective temperature, including thermal electron conduction and 
self-irradiation by photons from the compact object, assuming a 
NS with a pure hydrogen atmosphere and a magnetic field of 
less than 10$^{\rm 9}$ G. The fitting parameters 
of NSATMOS include the unredshifted effective temperature of the NS, 
$T_{\rm eff}$, the NS gravitational mass, $M_{\rm ns}$, the NS 
radius, $R_{\rm ns}$, the distance to the source, $D$, and the 
fraction of the neutron star surface that is emitting, $K$, which we fixed to 
1 throughout this work. We initially fixed the distance to 7.1 kpc and 
kept the NS mass and radius free. In order to compare the temperature 
evolution during the quiescent state of EXO 0748$-$676 with the 
results of \citet{degenaar2011}, 
we subsequently fixed the NS mass, radius and distance to 1.4 
M$_{\odot}$, 15.6 km and 7.4 kpc, respectively. The temperature 
for different instruments was linked to be the same value in the 
same observation, but was allowed to change freely among different 
observations.\footnote{There are two separate, but consecutive, observations 
on June 17 2010 under the same observation ID (see Table \ref{table1}). Here we linked 
the effective temperature to be the same in both.} 
To fit the positive residuals at around 0.5 keV, 
we included a collisionally-ionized diffuse gas model {\sc vapec} \citep{smith2001} in the fit. 
The parameters of this model are the plasma temperature, linked to be the same value for 
different instruments in the same observation, but allowed to change freely among 
different observations, the abundance of corresponding elements relative to the 
solar abundance, the redshift to the source (fixed to 0 in this analysis) and a normalisation. 
We considered the effect of interstellar absorption by including the component 
{\sc phabs} in XSPEC using the abundance tables of \cite{wilms2000}  with the 
photo-electric absorption cross-sections from 
\cite{balucinska1992} and the He cross-section by \cite{yan1998}. 
We also tested our fit using the abundance tables from  \cite{anders1989} and 
the cross-sections from \cite{verner1996}, respectively. It turned out that 
the selection of cross-sections did not change the result at all, while the 
choice of interstellar abundance do have an influence on the result, we mention 
this when necessary. 
We linked the hydrogen column density along the line of sight, $N_{\rm H}$, 
across all observations but left it free to change. We included a 
multiplicative factor to account for the different efficiency between 
different instruments. This factor was fixed to 1 for all PN data, and 
was left free but kept the same for all the MOS1 and MOS2 data separately.

\section{Results}
\label{results}

\subsection{Spectra}
\label{spectra}

In previous analyses of the quiescent emission of EXO 0748$-$676, 
the spectrum was usually fitted with a thermal component plus a power 
law with a photon index of $\sim$1.0$-$1.7 \citep{degenaar2011,guobao2011,diaz2011,degenaar2014}. 
We therefore initially tried the same model, 
{\sc const*phabs*(nsatmos + powerlaw)} in XSPEC with all parameters free. 
However, the parameters could not be well constrained unless we fixed the 
distance to the source. Here we adopted the value D = 7.1 kpc 
given by \cite{galloway2008b}. 
We also fixed the power-law index to 1.7, the same value used 
by \cite{degenaar2011}, which yields a $\chi^2$ of 3739 for 3674 degrees of freedom. The power-law component is, however, not significantly required to fit the 
data in any of the observations; the 95\% confidence upper-limit of the power-law 
normalisation for the four observations is in the range 
$1.0-6.0 \times 10^{-6}$ photons cm$^{-2}$s$^{-1}$keV$^{-1}$ at 1 keV,
which translates into an upper-limit of the power-law 
flux of less than 1\% of the total unabsorbed flux. This result does not change if we let the power-law index to vary between different observations, therefore we do not include a power law
to the model. The best fit with the model {\sc const*phabs*nsatmos} yields a 
$\chi^2$ of 3774 
for 3681 degrees of freedom, and the best-fitting parameters are 
$N_{\rm H}=(5.52^{+0.25}_{-0.30})
\times 10^{20} \rm{cm}^{-2}$, $M_{\rm ns} = 2.00^{+0.07}_{-0.23} M_{\odot}$ and  
$R_{\rm ns} = 11.3\pm1.2$ km, respectively. 
(Unless otherwise stated, all errors correspond to the 90\% confidence interval for a single interesting parameter.)
The hydrogen column density is 
slightly lower than the range, $N_{\rm H} = (0.7-1.2)\times10^{21} \rm{cm}^{-2}$, 
found during outburst by \citet{sidoli2005}, but it is consistent with the value 
found by \cite{degenaar2011}, \cite{diaz2011} and \cite{guobao2011} in quiescence. 
The best-fitting temperatures for the 4 observations are 
165 $\pm$ 16 eV, 161 $\pm$ 15 eV, 
160 $\pm$ 15 eV, 154 $\pm$ 15 eV, respectively
\footnote{\label{note1}These are the temperatures of the neutron-star atmosphere, 
not the temperatures seen by an observer at infinity.}. 
If we remove the flaring period in the spectra, the fit yields a $\chi^2$ of 3120 for 3065 degrees of freedom, and the best-fitting hydrogen column density, neutron-star mass and radius are $N_{\rm H}=(5.53^{+0.34}_{-0.30})
\times 10^{20} \rm{cm}^{-2}$, $M_{\rm ns} = 1.99^{+0.09}_{-0.26} M_{\odot}$, $R_{\rm ns} = 11.4\pm1.4$ km, respectively. The best-fitting temperatures for the four observations are 165 $\pm$ 16 eV, 161 $\pm$ 15 eV, 160 $\pm$ 15 eV, 153 $\pm$ 15 eV, respectively. All parameters are consistent, but with slightly larger error bars, compared with those obtained when we do not exclude the flares. 
The best-fitting parameters do not change significantly if we also add the RGS spectra (in the range $0.3-1.8$ keV) to the fits.

\begin{table*}
\caption{Fitting results to the spectra of EXO 0748$-$676}
\label{table2}
\begin{tabular}{ c c c c c c c c c}
\hline
\hline
Observation ID & MJD & \multicolumn{2}{c}
{NS $k T^{\infty}_{\rm eff}$ (eV)} & $kT_{\rm gas}$ & $F_{\rm X}$ 
& $F_{\rm bol}^{\rm th}$  & {\sc vapec} fraction\\
& & without {\sc vapec} & with {\sc vapec} & (eV) & ($10^{-12} \rm{erg~cm^{-2} 
s^{-1}}$) & ($10^{-13} \rm{erg~cm^{-2}s^{-1}}$) & (\%) \\
\hline
0605560401 & 54908 & 112.1$\pm$0.3 & 112.7$\pm$0.3 & 241$\pm$34 & 
9.45 $\pm$ 0.19 & 9.97 $\pm$ 0.21 & 7.4$\pm$1.8 \\
0605560501 & 55013 & 108.8$\pm$0.2 & 109.5$\pm$0.3 & 164$\pm$13 & 
8.45 $\pm$ 0.17 & 8.78 $\pm$ 0.16 & 9.8$\pm$1.5 \\
0651690101 & 55364 & 108.1$\pm$0.2 & 108.6$\pm$0.2 & 231$\pm$28 & 
8.12 $\pm$ 0.16 & 8.49 $\pm$ 0.16 & 6.9$\pm$1.5 \\
0690330101 & 56397 & 104.3$\pm$0.2 & 104.9$\pm$0.3 & 141$\pm$13 & 
7.05 $\pm$ 0.15 & 7.31 $\pm$ 0.15 & 10.0$\pm$1.7 \\
\hline
$\chi^{\rm 2}$
/d.o.f. & & 3852/3683 & 3657/3677\\
\hline\\
\end{tabular}
\\
\leftskip=0pt \rightskip=0pt {Note. Best-fitting results for
the model {\sc const*phabs*(nsatmos+vapec)} fixing the NS mass, radius 
and distance to 1.4 $M_{\odot}$, 15.6 km, and 7.4 kpc respectively. The 
best-fitting equivalent hydrogen column density 
$N_{\rm H}=(9.1\pm0.5)\times10^{20}~\rm{cm}^{-2}$, and the Nitrogen abundance of the hot gas component is 24$_{-6}^{+11}$ times solar abundance. 
The two columns under NS $k T_{\rm eff}^{\infty}$ show the NS effective
temperature at infinity for the model with and without the hot gas component, 
respectively. The column $kT_{\rm gas}$ shows the corresponding temperature of the hot gas component in each observation. 
$F_{\rm X}$ is the unabsorbed model flux 
in the $0.3-10.0$ keV band, and $F_{\rm bol}^{th}$ is the $0.01-100.0$ keV 
unabsorbed flux of the {\sc nsatmos} component. 
{\sc vapec} fraction is the fractional contribution of the 
hot gas component to the total unabsorbed flux in the $0.3-10.0$ keV energy range.}
\end{table*}

All the fits, however, show positive residuals at around 0.5 keV 
(see bottom panel of Figure \ref{spectra1}). 
We therefore added a hot gas component, {\sc vapec} in the analysis. 
The temperature 
and normalization of this collisionally-ionized plasma are linked to be the 
same for different instruments in the same observation but allowed to 
change freely among different observations. We initially set the gas abundance of C, N and O free at first. The fitting 
result yields an oxygen abundance $Z_{\rm O} = 1.1_{-0.5}^{+2.8}$, consistent with solar abundance within errors. The fit is insensitive to the carbon 
abundance; the best-fitting result gives 
$\chi^2=3620.2$ with $Z_{\rm C} = 0.007$ 
compared to $\chi^2=3620.6$ when $Z_{\rm C}=$1.  
Hence we fixed the carbon and oxygen abundance to solar abundance 
in the rest of the analysis. 
Since the normalisations of the {\sc vapec} component are consistent within errors for different observations, 
we linked them to be the same in all observations.
The best fit with the model {\sc const*phabs*(nsatmos+vapec)} gives a $\chi^2$ 
of 3620 for 3675 degrees of freedom. In this case, $N_{\rm H}$ is 
$(7.8\pm0.5)\times10^{20} \rm{cm}^{-2}$, and the temperature for the 4 
observations are 
167 $\pm$ 28 eV, 162 $\pm$ 28 eV, 160 $\pm$ 26 eV, 155 $\pm$ 27 eV, 
respectively (see footnote \ref{note1}). The best-fitting 
neutron star mass and radius are $M_{\rm ns} = 2.05 ^{+0.09} _{-0.39} M_{\odot}$ and $R_{\rm ns} = 11.4 \pm 2.1$ km, respectively. Again, the best-fitting
parameters do not change significantly if we also add the RGS spectra to the fits. 
The best-fitting Nitrogen abundance is 21$\pm$8 solar abundance, and the temperature 
of the hot gas for the 4 observations is 269$\pm$43 eV, 172$\pm13$ eV, 
250$\pm$29 eV and 152$\pm$16 eV, respectively. 
The F-test probability for a chance improvement when adding this collisionally ionised gas component 
to the model is 2$\times \rm{10^{-30}}$, which 
indicates that the addition of this component significantly improves the fit. 
Since the applicability of the F-test in these cases is questionable \citep{protassov2002}, 
to check this we simulated 10$^{4}$ spectra of the model without this hot gas component and fitted these spectra with the model that includes {\sc vapec}. None of these simulated spectra showed a normalisation as strong (or stronger) than the one we found from the fits to the data, which shows that the probability that this hot gas component is due to a statistical fluctuation is less than 10$^{-4}$. 
We did not find any significant edge in the effective area of the 
detectors around this energy that could, due to calibration uncertainties, 
account for these residuals. 

We explored whether the residuals at around 0.5 keV could be due to the cross-section 
and abundance tables used in the model that fits the interstellar absorption. To test this, we 
fitted the spectra with the model without the hot gas component using the cross-section table of \citet{verner1996}
and the abundances of \cite{anders1989}, but the residuals did not disappear, and the hot gas component was still 
significantly required by the fits. We also tried replacing the component {\sc phabs} by either {\sc 
vphabs} or {\sc tbnew} \citep[the newest version of {\sc tbabs}][]{wilms2000}, which allow us to change the abundance of the individual elements separately. One at
a time, we let the abundance of N, O, Ne and Fe free, but the residuals remained, and in all
cases the best-fitting abundances became either very low or very high. Finally, we added an extra edge to the model to check whether the line could be a calibration artefact related to the oxygen edge in the detectors. We fixed the energy of this edge in the model to 0.538 keV, the energy of the OI edge, and we allowed the normalisation of the edge to be either positive or negative to account for, respectively, a lower or higher amount of oxygen contamination in the detectors relative to the values in the XMM-Newton calibration files. Since the positive residuals in our fits appear at somewhat lower energy than that of the OI edge, this procedure did not improve the fit. In summary,
none of these alternatives could explain the residuals, and we therefore continued 
including this additional component in our model.

\begin{figure}
\begin{center}
\includegraphics[angle=270,width=8cm]{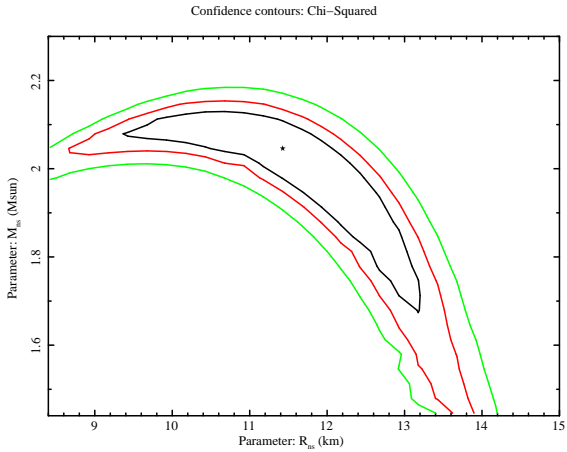}
\caption{Contour plot of the mass and radius parameters of the neutron star in EXO 0748$-$676. 
We fitted the model {\sc const*phabs*(nsatmos+vapec)} in the energy range $0.3-10.0$ keV 
with the distance to the source fixed to 7.1 kpc. The cross marks the best-fitting mass and radius, 
$M_{\rm ns} = 2.05~M_{\odot}$ and $R_{\rm ns} = 11.4$ km, respectively.
The three contour lines represent the confidence levels of 68 per cent (black), 
90 per cent (red) and 99 per cent (green) for two parameters.}
\label{mr1}
\end{center}
\end{figure}

In Figure \ref{spectra1} we show the spectra and the best-fitting model 
with (middle panel) and without (bottom panel) the hot gas component, 
and in Figure \ref{mr1} we show the contour plot of the mass and radius 
obtained from the fits.

\begin{figure*}
\begin{center}
\subfigure[exponential decay]{\includegraphics[width=8cm]{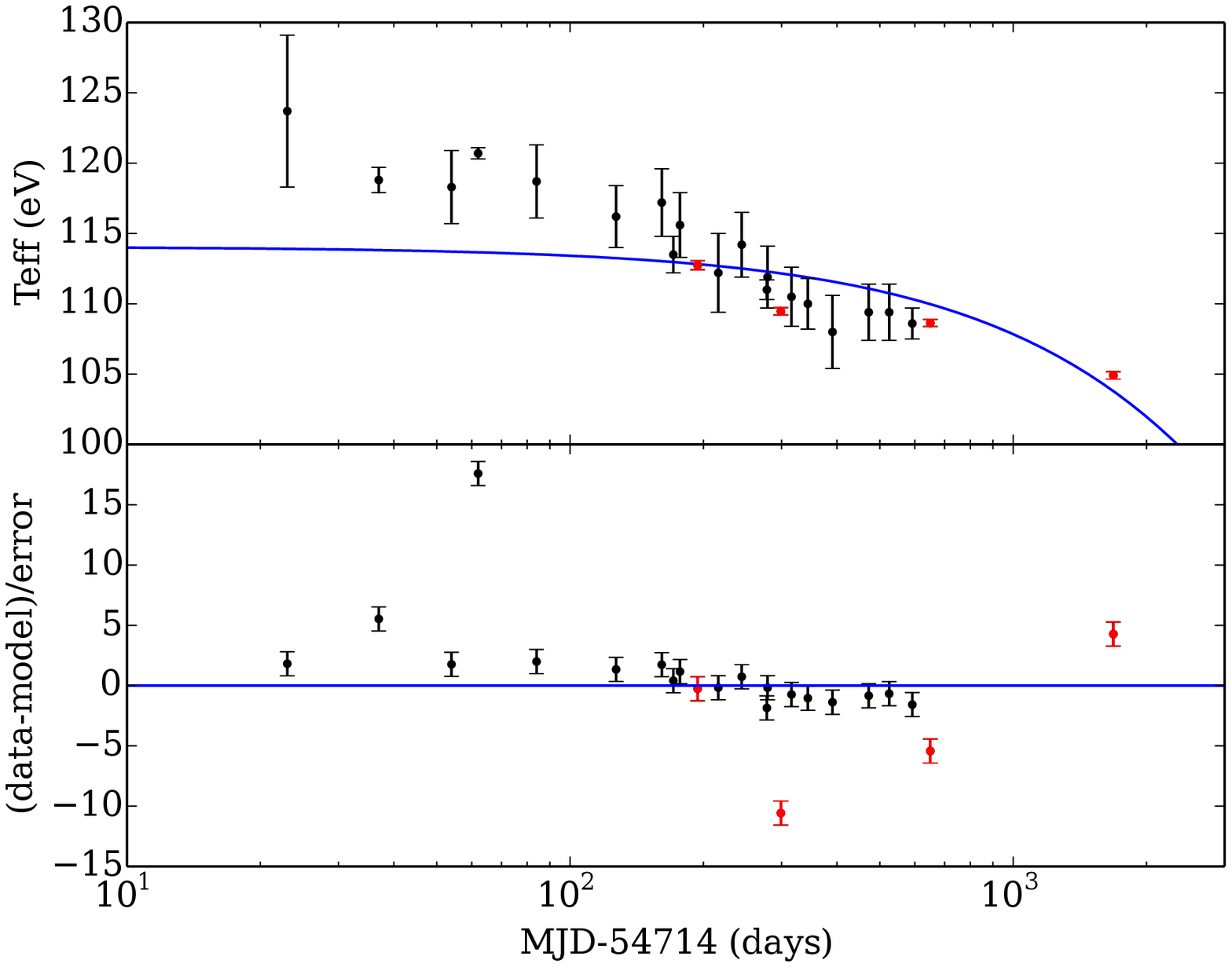}}
\subfigure[exponential decay plus a constant]{\includegraphics[width=8cm]{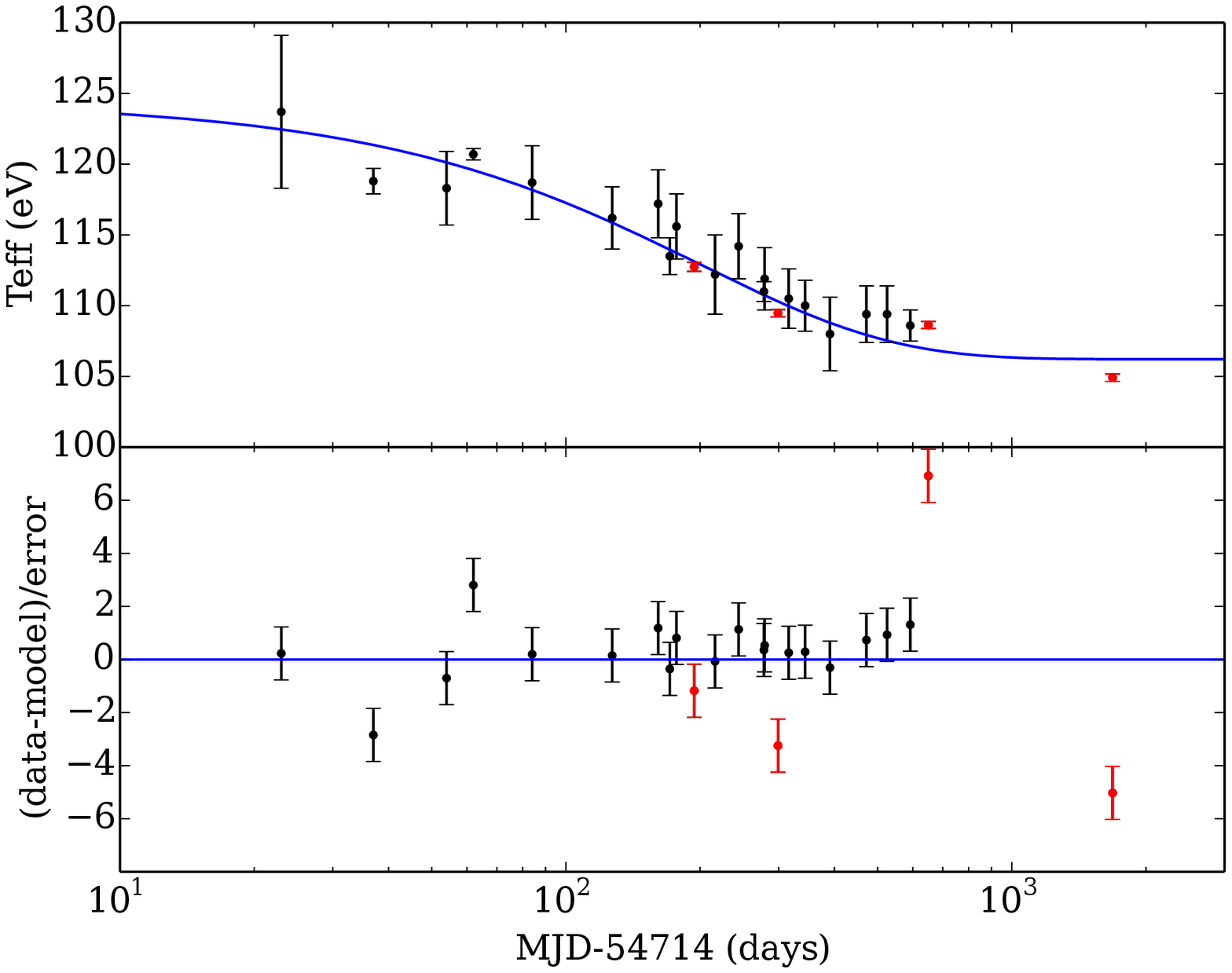}}
\subfigure[power-law function]{\includegraphics[width=8cm]{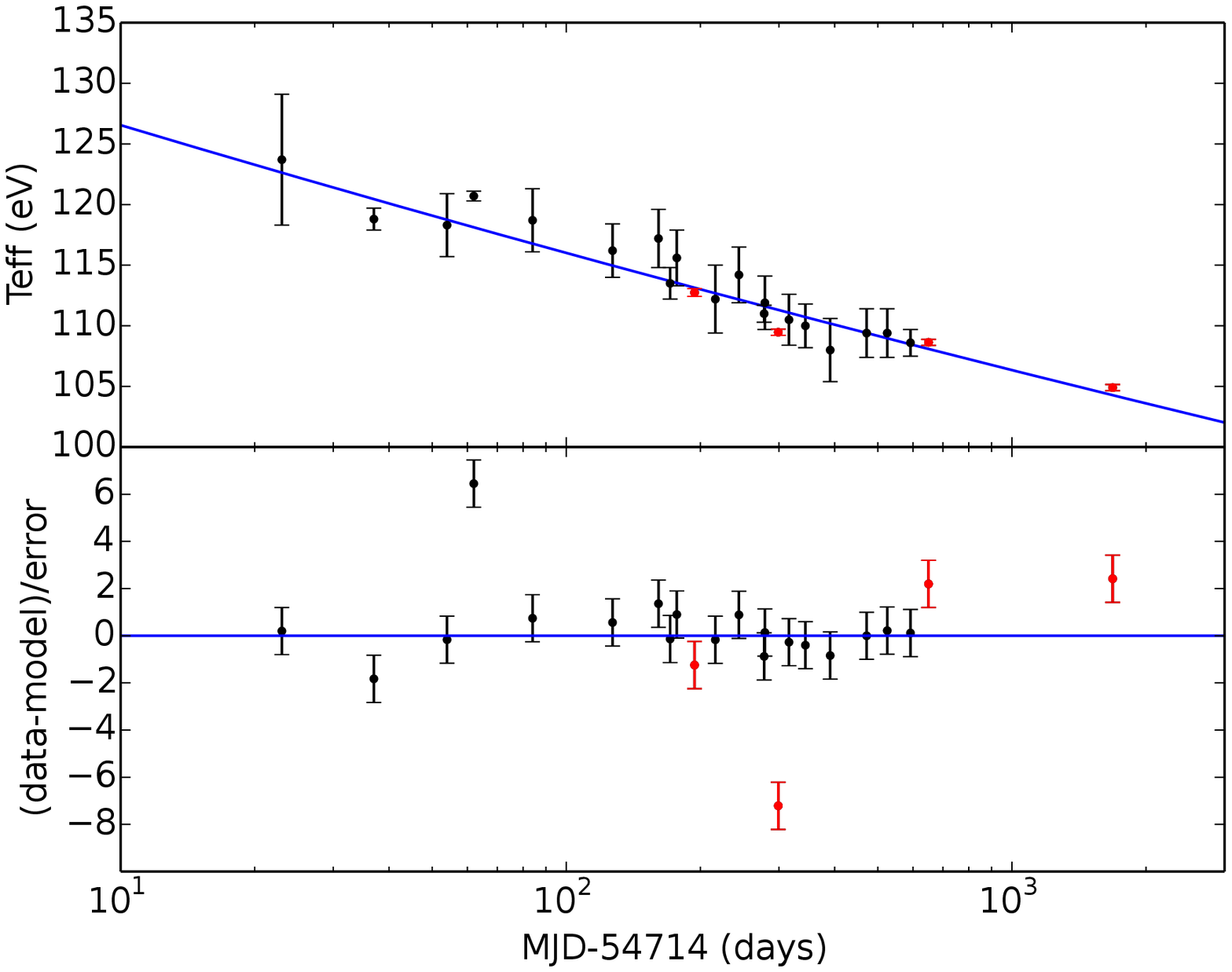}}
\subfigure[broken power-law]{\includegraphics[width=8cm]{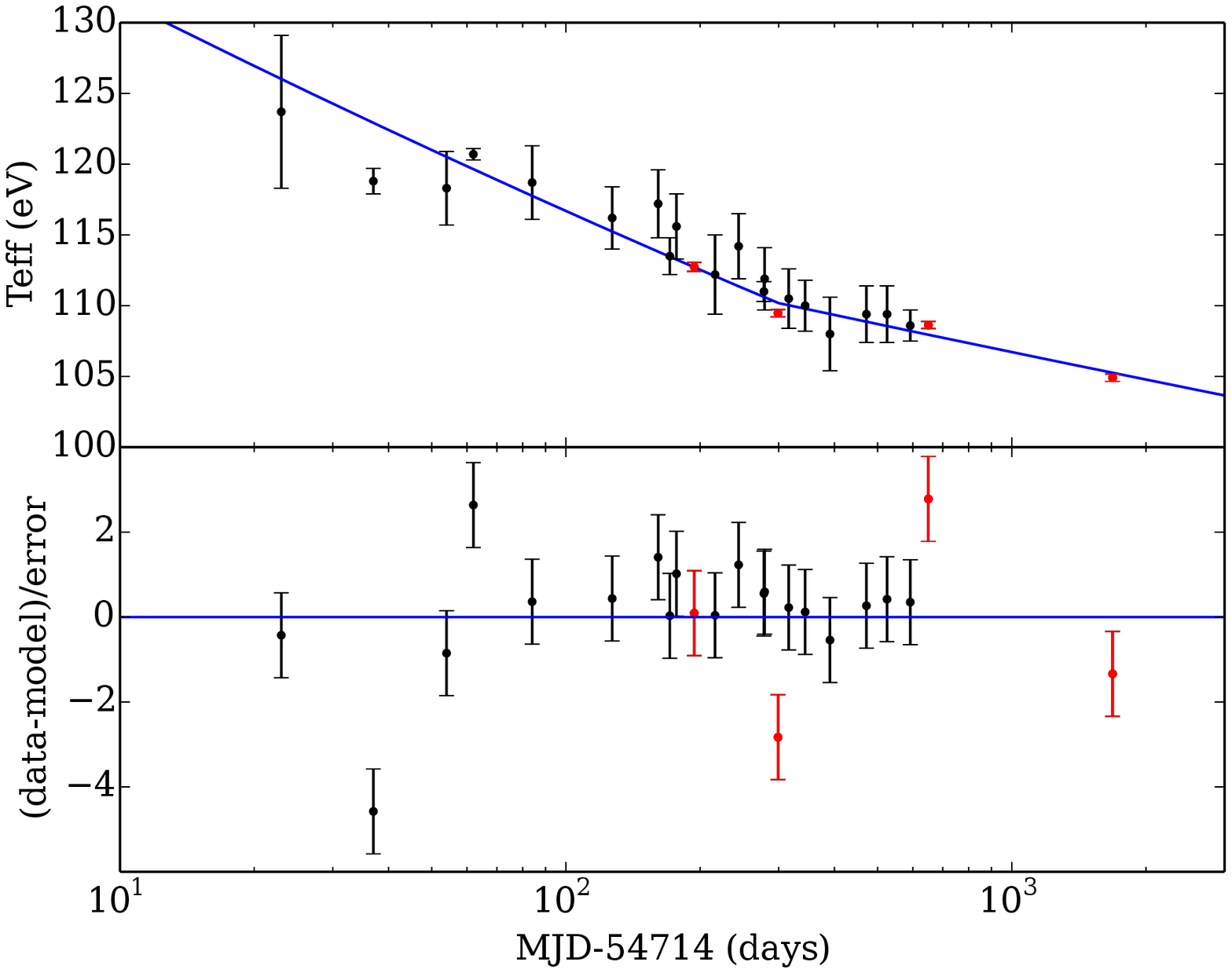}}
\caption{Effective temperature evolution of EXO 0748$-$676 fitted with 
different functions: exponential decay (upper left panel), exponential decay
plus a constant offset (upper right panel), power-law decay (lower left panel)
and broken power-law decay (lower right panel). 
The red circles represent the data obtained from the XMM-Newton observations in 
this paper; the black circles show the Chandra, Swift data and the first XMM-Newton observation presented here
that were used in the analysis of \protect\cite{degenaar2011}.
The start date MJD 54714, corresponding to 2008 
September 5, is the same date chosen by \protect\cite{degenaar2011}
}
\label{coolingcurve}
\end{center}
\end{figure*}

In order to compare the temperature of the neutron star in EXO 0748$-$676
in our XMM-Newton observations with the temperature in the Chandra and Swift
observations in \cite{degenaar2011}, we fixed the mass and the radius of the
neutron star and the distance to the source to the same values used by
\cite{degenaar2011}, $M_{\rm ns}$=1.4 M$_{\odot}$, $R_{\rm ns}$=15.6 km, and 
$D$=7.4 kpc. In this case the best fitting model gives 
$N_{\rm H}=(9.1\pm0.5)\times10^{20} \rm{cm}^{-2}$ and 
the Nitrogen abundance of the hot gas component is 
$Z_{N\rm}$=24$_{-6}^{+11}$, 
with a $\chi^{2}$ of 3657 for 3677 degrees of freedom.
We then calculated the effective temperature of the NS seen by an observer 
at infinity, $kT_{\rm eff}^{\infty}$, using the formula $kT_{\rm eff}^{\infty}=kT_{\rm eff} 
(1-R_{\rm s}/R_{\rm NS})^{1/2}$, where $kT_{\rm eff}$ is the best-fitting temperature
in our models, $R_{\rm s}=2GM_{\rm NS}/c^{2}$, 
$G$ is the gravitational constant and $c$ is the speed of light.
The NS effective temperatures at infinity for the four 
observations are given in Table \ref{table2}.
We also fitted the spectra without the hot gas component to check 
whether this component had an effect on the best-fitting temperatures; 
in this case the best fit gives 
$N_{\rm H}=(6.7\pm0.2)\times10^{20} \rm{cm}^{-2}$
with a $\chi^2$ of 3852 
for 3683 degrees of freedom, and the 
NS effective temperatures at infinity for the four observations are 
given in Table \ref{table2}.

The average flux from the hot gas component is about 
$5.8 \times 10^{-14} \rm{erg~cm^{-2} s^{-1}}$, which is 
about 7$-$10\% of the total flux in the energy range from 0.3 to 10.0 keV. 
The inclusion of this component in the model does not change the temperatures of the NS significantly as shown in Table \ref{table2}.  

\subsection{Cooling curves}

\begin{table*}
\caption{Fitting results of the cooling curve of EXO 0748$-$676}
\label{table3}
\begin{tabular}{ l l }
\hline
\hline
\multicolumn{2}{c}{Exponential decay}\\
\multicolumn{2}{c}{$y(t) = a e^{-(t-t_{0})/b}$}\\
\hline
Normalization factor, $a$ & 114.1 $\pm$ 0.9 eV\\
e-folding time, $b$ & 17844 $\pm$ 3124 days\\
$\chi^{\rm 2}$ 
(d.o.f.) & 528 (21)\\
\hline
\multicolumn{2}{c}{Exponential decay with constant offset}\\
\multicolumn{2}{c}{$y(t) = a e^{-(t-t_{0})/b}+c$}\\
\hline
Normalization factor, $a$ & 18.2 $\pm$ 1.4 eV\\
e-folding time, $b$ & 200 $\pm$ 27 days\\
Constant offset, $c$ & 106.2 $\pm$ 0.5 eV\\
$\chi^{\rm 2}$ 
(d.o.f.) & 109 (20)\\
\hline
\multicolumn{2}{c}{Power law}\\
\multicolumn{2}{c}{$y(t) = a (t-t_{0})^{-b}$}\\
\hline
Normalization factor, $a$ & 138.0 $\pm$ 2.1 eV\\
Power law index, $b$ & 0.04 $\pm$ 0.01\\
$\chi^{\rm 2}$ 
(d.o.f.) & 115 (21)\\
\hline
\multicolumn{2}{c}{Broken power law}\\
\multicolumn{2}{c}{$y(t) = a (t-t_{0})^{-b}$,~$t-t_{0} \le t_{b}$;}\\
\multicolumn{2}{c}{$~~~~~~~~~~~~~~~~~~~~~~a (t-t_{0}-t_{b}) (t-t_{0})^{c}$,~$t-t_{0} > t_{b}$}\\
\hline
Normalization factor, $a$ & 148.5 $\pm$ 3.5 eV\\
Power law index $b$ & 0.05 $\pm$ 0.01\\
Power law index $c$ & 0.03 $\pm$ 0.01\\
Break point $t_{0}$ & 299 $\pm$ 73 days\\
$\chi^{\rm 2}$ 
(d.o.f.) & 53 (19)\\
\hline
\hline
\leftskip=0pt \rightskip=0pt {Note. All errors correspond
to the 1$\sigma$ confidence levels.}
\end{tabular}
\end{table*}

In our analysis we assumed the same NS mass, radius and source distance 
as \citet{degenaar2011}, which allows us to compare the effective 
temperature in our work with those in their work. Therefore, we combined 
our temperatures with those obtained by \cite{degenaar2011} from Chandra, 
Swift and previous XMM-Newton observations. 

In order to investigate the temperature evolution, we first fit 
the effective temperature vs. time with an exponential function, 
$T_{\rm eff} = a e^{-(t-t_{0})/b}$; the fit yields
$a = 114.1 \pm 0.9$ eV with an e-folding time of 
$17844 \pm 3124$ days. (The start time of the cooling was fixed at  
${t_{\rm 0} = 54714}$ MJD days as in \citealt{degenaar2011}.) The result of this 
fit is shown in the upper left panel of Figure \ref{coolingcurve}. 
The fit is not good, with a reduced $\chi^{2}$, 
$\chi^{\scriptscriptstyle 2}_{\scriptscriptstyle \nu}$, of 25 for 21 degrees 
of freedom. We then introduced an additional constant offset $c$ 
in the exponential function. This model gives a better fit 
($\chi^{\scriptscriptstyle 2}_{\scriptscriptstyle \nu} = 6$ for 
20 degrees of freedom) with  
$a = 18.2 \pm 1.4$ eV, $b = 200 \pm 27$ days, and the constant offset
$c = 106.2 \pm 0.5$ eV. The result is shown in the upper right panel of 
Figure \ref{coolingcurve}. We also fitted a power-law and a broken 
power-law to the data, as proposed by \cite{brown2009}.
For the power-law function, $y(t) = a (t-t_{\rm 0})^{-b}$, we get 
$a = 138.0 \pm 2.1$ eV and $b = 0.04\pm 0.01$. 
The result is shown in the 
lower left panel of Figure \ref{coolingcurve}. The lower right panel of 
Figure \ref{coolingcurve} shows the fit of a broken power-law. The best fit 
gives a normalization factor $a = 148.5 \pm 3.5$ eV with a break at 
$299 \pm 73$ days. The power law index before and after the break point 
are $0.05 \pm 0.01$ and $0.03 \pm 0.01$, respectively.
All the fitting parameters can be found in Table \ref{table3}.

\cite{degenaar2014} re-analysed the Chandra data of EXO 0748$-$676 used in \cite{degenaar2011}, including also the three XMM-Newton observations available at that time. They fitted these observations simultaneously, fixing the NS mass to $1.64~M_{\odot}$ and the distance to the source to 7.1 kpc, which yielded a best-fitting NS radius of $13.2^{+0.6}_{-2.0}$ km. Because the NS mass and radius used in the fits by \cite{degenaar2014} are different from those in 
\cite{degenaar2011} that we used for Figure 3, we cannot compare the temperatures in \cite{degenaar2014} and those shown here (Figure 3). We therefore re-fitted our spectra fixing the mass and radius to the values in \cite{degenaar2014}, and verified that the best fitting parameters to the cooling curves that include the results of \cite{degenaar2014} and ours are consistent with those in Table 3.

\section{Discussion}
\label{discussion}

We analysed four XMM-Newton observations, from 2009 to 2013, of the neutron-star low-mass X-ray binary EXO 0748$-$676 in quiescence. We fitted the X-ray spectra in the $0.3-10$ keV range with the neutron-star model atmosphere NSATMOS. If we fix the distance to 7.1 kpc (Galloway et al. 2008), the fit yields a $\chi^2$ of 3774 for 3681 degrees of freedom, and the best-fitting mass and radius of the neutron star are, respectively, 
$M_{\rm ns} = 2.00^{+0.07}_{-0.23}~M_{\odot}$ and $R_{\rm ns} = 11.3\pm1.2$ km. 
Although this fit is statistically acceptable, we get a significantly better fit ($\chi^2=$3620 for 3675 dgrees of freedom ) with consistent values of the neutron-star mass and radius,
$M_{\rm ns} = 2.05^{+0.09}_{-0.39}~M_{\odot}$ and 
$R_{\rm ns} = 11.4 \pm 2.1$ km, 
if we add a hot gas component to the model contributing 7$-$10\% of the total 
unabsorbed flux of the source in the $0.3-10.0$ keV range. 
Combining the neutron-star temperatures from our fits with those obtained by \cite{degenaar2011} from Chandra and Swift observations, we find that the cooling curve of the neutron-star in EXO 0748$-$676 is compatible with a model consisting either of an exponential decay plus a constant or a (broken) power law, although formally none of the models gives a good fit. We further found that no extra emission from a high-energy component, usually represented by a power law in the model, is required to fit the spectra of these four XMM-Newton observations, with a 95 per cent confidence upper limit  of 1 per cent to the contribution of the power law to the total flux of the source in the 0.5$-$10.0 keV range.

We should mention that the choice of the table of solar abundance in the ISM and the 
distance to the source both significantly affect the best-fitting neutron-star 
mass and radius. 
If we use the abundance table of \cite{anders1989} instead of the one of \cite{wilms2000}, 
the best-fitting mass and radius given by the neutron star hydrogen atmosphere model 
are 2.17$^{+0.06}_{-0.13} M_{\odot}$ and 11.8$^{+0.9}_{-1.2}$ km, respectively. 
The value of the distance we used in our analysis was obtained 
by \citet[][see also \citealt{galloway2008b}]{galloway2008a} from three photospheric 
radius expansion X-ray bursts of the source under the assumption 
of a canonical neutron-star mass $M_{\rm ns} = 1.4~M_{\odot}$ and 
radius $R_{\rm ns} = 10~$km. 
In fact, the determination of the distance not only depends upon the 
neutron-star mass and radius, 
but also upon the hydrogen mass fraction of the burning fuel of the bursts. 
Assuming different hydrogen abundances in the burst fuel yields 
different values of the mass and radius. 
In addition, the angular distribution of the 
burst radiation, which was neglected by \cite{galloway2008b}, 
plays an important role in determining the source distance \citep{lapidus1985}, and affects the best-fitting mass and 
radius significantly. The lowest energy of the spectra used in the analysis also has a strong impact on the best-fitting neutron star mass and radius 
in EXO 0748$-$676. 
We expand further on these issues in \S4.3, and we will discuss this effect in more detail in a separate paper. 

\subsection{Residual emission near 0.5 keV}
\label{line}

There is no significant edge in the effective area of the XMM-Newton instruments at this energy, which makes it unlikely that the residuals around 0.5 keV are instrumental. The residuals are still significant when we either use different cross-section and abundance tables or let the abundance of N, O, Ne and Fe in the interstellar medium free. 
The fits with the model {\sc const*phabs*nsatmos} show significant residuals at $\sim$0.5 keV (see Fig. 1). 
To fit the residuals at around 0.5 keV, we included a hot gas component, {\sc vapec}, in 
the model. The addition of this component improves the fits significantly; 
this component contributes $\sim7-10$\% of the total flux in the 0.3$-$10.0 keV range. 

This hot collisionally ionised plasma could correspond to the absorption 
component reported by \cite{vanpeet2009}
. They analysed the spectra of XMM-Newton observations during the outburst phase of 
this source and, by comparing the dipping and persistent spectra, they found that there are two absorbers in this system, one photoionised and one collisionally ionised. The latter should be located sufficiently far from the central source; 
for an assumed distance of $\sim10^{11} \rm{cm}$ away from the neutron star this gas 
would have a particle density $n>10^{14} \rm{cm}^{-3}$, and does not change from persistent to dipping states. From our fits with {\sc vapec}, 
and assuming a geometry similar to that in \cite{vanpeet2009}, we find a particle 
density of n$\sim10^{13} \rm{cm}^{-3}$. The temperature of this collisionally 
ionised plasma is around 150$-$270 eV, whereas \cite{vanpeet2009} found a 
temperature between 60$-$80 eV. 
This difference could be due to the higher luminosity of the source in 
\cite{vanpeet2009} compared to our observations, as the radiation from 
accretion during the outburst could cool down this plasma. 
This is also consistent with the findings of \cite{vanpeet2009}, that the 
column density increases significantly while this gas component is less ionised during 
the dips, as the material coming from the accretion stream cools down the plasma. 

The residuals at $\sim$ 0.5 keV can also be fitted by adding a Gaussian component 
to the model as {\sc const*phabs*(nsatmos+gaussian)}. This model gives a 
moderately broad emission line with $E = 0.48\pm0.03$ keV and 
$\sigma$ = 0.09$\pm0.03$ keV, with a $\chi^{2}$ of 2627 for 3675 
degrees of freedom ($N_{\rm H}$=(7.9$\pm0.9)\times10^{20} \rm{cm^{-2}}$, 
$M_{\rm ns}=2.14_{-0.26}^{+0.07} M_{\odot}$, $R_{\rm ns}$=11.4$\pm$1.6 km). 
The F-test probability for a chance improvement when adding a line to the model 
is 5$\times \rm{10^{-29}}$, which indicates that the addition of the Gaussian 
significantly improves the fit. 
As in the case of the vapec component, to check this we simulated 10$^{5}$ 
spectra of the model without the Gaussian line and fitted these spectra with the 
model that includes the line. None of these simulated spectra showed a line as 
strong (or stronger) than the one we found from the fits to the data, which shows 
that the probability that the line is due to a statistical fluctuation is less 
than 10$^{-5}$. 

The energy of the line, $E = 0.48\pm0.03$ keV, is consistent with the Ly$\alpha$ transition of NVII (rest energy of 0.500 keV).
\cite{maitra2011} reported a strong emission line near 0.5 keV in the spectrum of the transient low-mass X-ray binary MAXI J0556$-$332 in outburst. \cite{maitra2011} concluded that the line in MAXI J0556$-$332 is due to NVII, and likely originates in material accreting from a donor star with an unusually high O/N abundance. Using the RGS spectrometer on board XMM-Newton, \cite{cottam2001} found several emission lines, among them the NVII Ly$\alpha$ line, in the spectrum of EXO 0748$-$676 in outburst. \cite{pearson2006} found a strong, $16.5$\AA\ equivalent width, NV $\lambda1240$ line in the HST spectrum, and \cite{mikles2012} identified NIII lines in the optical spectrum of EXO 0748$-$676 in outburst. All these results indicate a significant nitrogen abundance in the companion star, UY Vol, supporting the identification of the line.


If the line is indeed from the source, and the NVII identification is correct, given the low luminosity of the source the highly ionised nitrogen must be produced by collisional ionisation in a plasma at $\sim 0.1$ keV or hotter \citep{cox1969}. This high temperature rules out the surface of the companion star as the place where the line is formed, while the lack of a significant gravitational redshift rules out the neutron-star atmosphere. 
On the other hand, the moderate width of the line in our fits suggests that the line may be produced in a residual accretion disc, or an advection dominated accretion flows (ADAF, \citealt{narayan1994}; but see \citealt{menou2001}) around the neutron star in quiescence. If this is indeed the case, the quiescent spectrum of EXO 0748$-$676 (and possibly other sources) may contain not only emission from the neutron-star surface, but also from a weak accretion disc, which would significantly affect the study of cooling neutron stars in binary systems. Alternatively, if a residual disc is only present in EXO 0748$-$676, this could be the reason why the NS cooling process in this source appears to be less efficient than in the case of the NS in KS 1731--260, MXB 1659$-$29 and XTE J1701$-$462 \citep[][see also \S\ref{cooling} below]{degenaar2011,diaz2011,homan2014}.

\subsection{The cooling of the neutron star in EXO 0748--676}
\label{cooling}
After being heated by accretion during the outburst period, in the quiescent phase the crust of neutron stars in LMXBs takes several years to cool down and regain thermal equilibrium with the NS core \citep{yakovlev2004}. In Figure \ref{coolingcurve} we show the evolution of the neutron-star temperature in EXO 0748$-$676 over a period of 5 years in quiescence, after the source had been active for more than 20 years. In this Figure we included the data of the XMM-Newton observations analysed here, plus the temperatures obtained by \cite{degenaar2011} using Chandra and Swift. 

The cooling curve cannot be fitted with a simple exponential decay, whereas an exponential decay with a constant offset, a power-law or a broken power-law decay all fit the data reasonable well. In the case of an exponential decay plus a constant the e-folding time is $200 \pm 27$ days, and the constant temperature level is $106.2 \pm 0.5$ eV. 
The decay time provides the thermal relaxation time of the NS crust \citep{brown2009}. 
Using Chandra and Swift data, \cite{degenaar2014} found an e-folding time of $172 \pm 52$ days and a constant temperature level of $114.4 \pm 1.2$ eV while, based on the same first three observations presented here plus another XMM-Newton from 2008, \cite{diaz2011} found an e-folding time of $133.5 \pm 87.8$ days and a constant base temperature of $109.1 \pm 2.2$ eV. Given that the neutron-star temperature in the last observation of EXO 0748$-$676 presented here is already $\sim 105$ eV, if it is the case, then the NS crust must already be close to the equilibrium temperature of the core. 
The three Chandra observations \citep{degenaar2014} between our last two XMM-Newton observations are consistent with this scenario. By fitting a blackbody model to a serendipitous Einstein IPC observation of EXO 0748$-$676 in quiescence taken in 1980, before it was discovered as a bright transient, \cite{garcia1999} obtained a pre-outburst temperature of $220^{+140}_{-100}$ eV. Fitting an {\sc nsatmos} model simultaneously to the same pre-outburst Einstein and the post-outburst Chandra and XMM-Newton spectra, \cite{degenaar2014} found a pre-outburst temperature of $94^{+5.6}_{-16.0}$ eV. If the latter is correct, a further decrease in temperature may be expected.

The cooling of EXO 0748$-$676 appears to be less pronounced than that of other sources \citep{degenaar2014,homan2014}. According to the numerical simulations of \citealt{brown2009}, the NS cooling in these systems proceeds as a broken power law, with the initial power law being directly related to the thermal energy accumulated in the outer NS crust during outburst. \citet[][see also \citealt{diaz2011}]{degenaar2011} noted that the effective temperature of the other similar sources degreased by $20 - 40$ percent. \cite{degenaar2014} discussed the possibility that the stalled cooling of EXO 0748$-$676 one year after the source entered quiescence may be due to either heat driven by convection in the outer layers of the NS \citep{medin2011,medin2014}, or to the temperature profile in the crust not having reached a steady state at the end of the outburst. The latter mechanism would be applicable to sources that display short ($\sim 1$ year) and very bright ($L \sim L_{\rm Edd}$) outbursts rather than to the case of EXO 0748$-$676, which was active for at least 24 years accreting at less than $\sim 5\%$ the Eddington rate \citep{homan2003}. On the other hand, \cite{schatz2014} identified a neutrino cooling mechanism in the NS crustal shell that is very sensitive to temperature. It may be possible that this mechanism plays a role in this case, since EXO 0748$-$676 is one of the two hottest crust cooling NS \citep{degenaar2014}. 

Our results suggest another possibility for the slow cooling of EXO 0748$-$676: If the emission line in the spectra of all our observations (see \S4.1) is from the source, it suggests that EXO 0748$-$676 may still be experiencing low-level accretion from a residual accretion disc in the quiescent phase. If this is correct, the emission in all observations of EXO 0748$-$676 in quiescence may actually be the combination of thermal emission from the neutron star and from a relatively cool accretion disc. 
From the width of the $\sim0.5$-keV Gaussian line, and assuming that the line is produced close to the inner edge of this putative disc, the inner radius of the disc would be $\sim 30-100$ km for a neutron-star mass in the range $1.4 - 2~M_\odot$ and, for disc temperatures of the order of 100 eV (\S\ref{line}), the disc would contribute $\sim 5-10$ per cent of the $0.3-10$ keV flux, and would therefore contaminate the thermal spectrum of the cooling neutron star. If this is the case, then this disc component may bring additional uncertainties in the neutron star mass and radius analysis.

\citet{hynes2009} detected a modulation in the quiescent optical and infrared light curves of EXO 0748$-$676 between November 2008 and January 2009, with a period consistent with the X-ray orbital period of the source. They concluded that even in quiescence either the accretion disc or emission from the X-ray heated inner face of the companion star dominates the optical/infrared emission of this system. \citet{Bassa2009} found a broad component in the emission lines in a Doppler tomography study of EXO 0748$-$676 in the early phases of quiescence, which they interpreted as emission from a weak accretion disc. On the other hand, using observations later on in the quiescent state of EXO 0748$-$676, \cite{Ratti2012} did not detect the broad component reported by \citet{Bassa2009}, and concluded that, at least at that time, there was no significant contribution from the accretion disc to the optical emission. 

\subsection{The power law}

\cite{degenaar2009} found that in a 2008 Chandra observation, just after EXO 0748$-$676 had turned into quiescence, besides a thermal component,
a power-law component with an index of 1 contributed $\sim16 - 17$ per cent of the total $0.5-10$ keV flux. A month after this Chandra observation, using an XMM-Newton observation \cite[][see also \citealt{diaz2011}]{guobao2011} found that a power-law component contributing $\sim 10$ per cent of the total flux in the same energy band was required to fit the spectra. Furthermore, using Chandra data, \cite{degenaar2011} found that the power-law component in their fits changed irregularly, decreasing significantly from $20 \pm 3$ per cent in October 2008 to $4 \pm 3$ per cent in June 2009, and increasing again to $15 \pm 4$ per cent in April 2010. A similar behaviour has also been observed in several NS X-ray binaries in quiescence, which has been interpreted as possible low-level accretion in these systems \citep{cackett2010,cackett2010a,fridriksson2010}.

EXO 0748$-$676 was observed five times in quiescence with XMM-Newton. The spectrum of the first observation, done almost immediately after accretion switched off, showed a thermal and a power-law component. 
Here we find that the spectra of the other four observations between 2009 to 2013, can be well described only by a thermal component (here we used the NS atmosphere model {\sc nsatmos}) and an emission line at $\sim 0.5$ keV (see \S\ref{spectra}). Our fits do not require any power-law component, with a 95\% confidence upper limit of 1\% to the possible contribution of the power law to the 0.5$-$10.0 keV unabsorbed flux of the source. This upper limit is similar to those found by \citet{diaz2011} in their fits of the first three observations presented here, and implies that, at least in EXO 0748$-$676, any possible accretion onto the magnetosphere or a shock from a pulsar wind \citep{campana1998} becomes negligible as the cooling process of the neutron star surface proceeds. 

\subsection{The mass and radius of the neutron star in EXO 0748--676}
\begin{figure*}
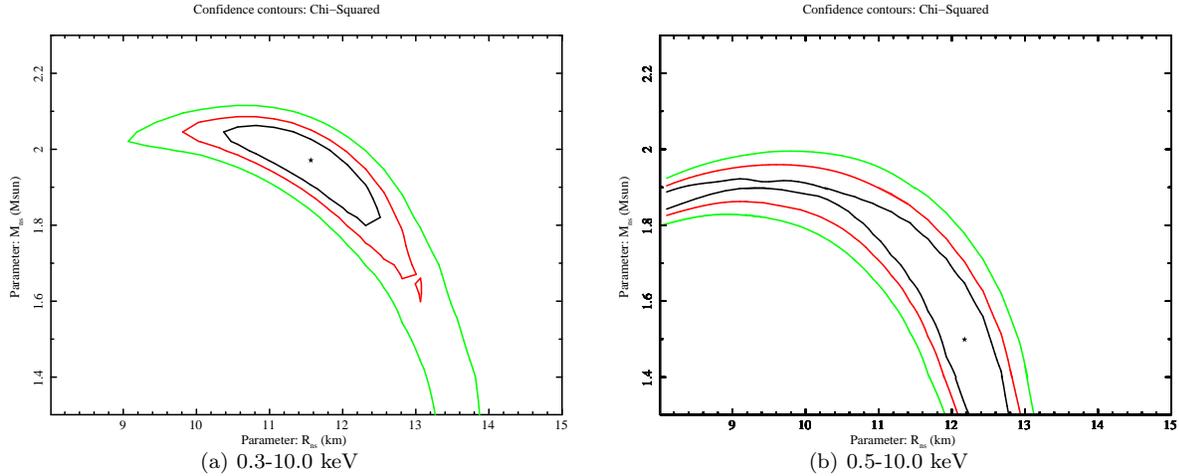

\begin{center}
\subfigure[0.3-10.0 keV]{\includegraphics[angle=270,width=8cm]{fit4_contour1_thermal_3e-1_2b}}
\subfigure[0.5-10.0 keV]{\includegraphics[angle=270,width=8cm]{fit4_contour1_thermal_5e-1_2d}}
\caption{Contour plots of the mass-radius relation of the neutron star in 
EXO 0748$-$676 based on the fits to the spectra in different energy ranges. 
We used the model {\sc const*phabs*nsatmos} with the distance fixed to 7.1 kpc. 
The left panel is for the fit in the $0.3-10.0$ keV energy range; the most likely 
mass and radius are $M_{\rm ns}=1.99$ M$_{\odot}$ and $R_{\rm ns}=11.3$ km, 
respectively. The right panel shows the result for the fit in the $0.5-10.0$ keV 
energy band; the most likely mass and radius are $M_{\rm ns}=1.50$ M$_{\odot}$ 
and $R_{\rm ns}=12.2$ km, respectively. 
The three contour lines represent the confidence level of 68 per cent (black), 90 per cent (red) and 99 per cent (green) for two parameters. } 
\label{mr2}
\end{center}
\end{figure*}

\cite{degenaar2014} found a NS mass $M_{\rm ns}$=${\rm 1.64 \pm 0.38}$ M$_{\odot}$ and a radius $R_{\rm ns}$=$\rm {13.2^{+0.6}_{-2.0}}$ km from fits to the quiescent spectra of EXO 0748$-$676 in the 0.5$-$10.0 keV band, whereas in our analysis we used the $0.3-10$ keV band of the EPIC cameras; furthermore, \cite{degenaar2014} did not include an additional component at around 0.5 keV in their model. 
Therefore, to study the influence of these factors upon the best-fitting mass, and to compare our results with those of \cite{degenaar2014}, we re-fitted our data both in the 0.3$-$10.0 keV and in the 0.5$-$10.0 keV band with the model {\sc const*phabs*nsatmos}.
In both cases we kept the distance to the source fixed at 7.1 kpc, similar to \cite{degenaar2014}. In the two panels of Figure \ref{mr2} we show the contour plots of the best-fitting NS mass and radius from this analysis. The left panel of this Figure shows the result of the fits in the 0.3$-$10.0 keV band; the most likely mass and radius are $M_{\rm ns}=2.08^{+0.07}_{-0.15}~M_{\odot}$ and $R_{\rm ns}=11.9\pm0.7$ km. As we mentioned in \S\ref{spectra}, these values are consistent with the ones for the model {\sc const*phabs*(nsatmos+vapec)}, which demonstrates that adding the hot gas component in the model does not have a significant effect in the best-fitting mass and radius of the NS. The right panel of Figure \ref{mr2} shows the result of the fits in the 0.5$-$10.0 keV band; in this case the most likely mass and radius are $M_{\rm ns} = 1.50 ^{+0.40}_{-0.99}$ M$_{\odot}$ and $R_{\rm ns} = 12.2 ^{+1.0}_{-3.6}$ km, similar to the values found by \cite{degenaar2014}. 

\begin{figure*}
\begin{center}
\subfigure{\includegraphics[width=8cm]{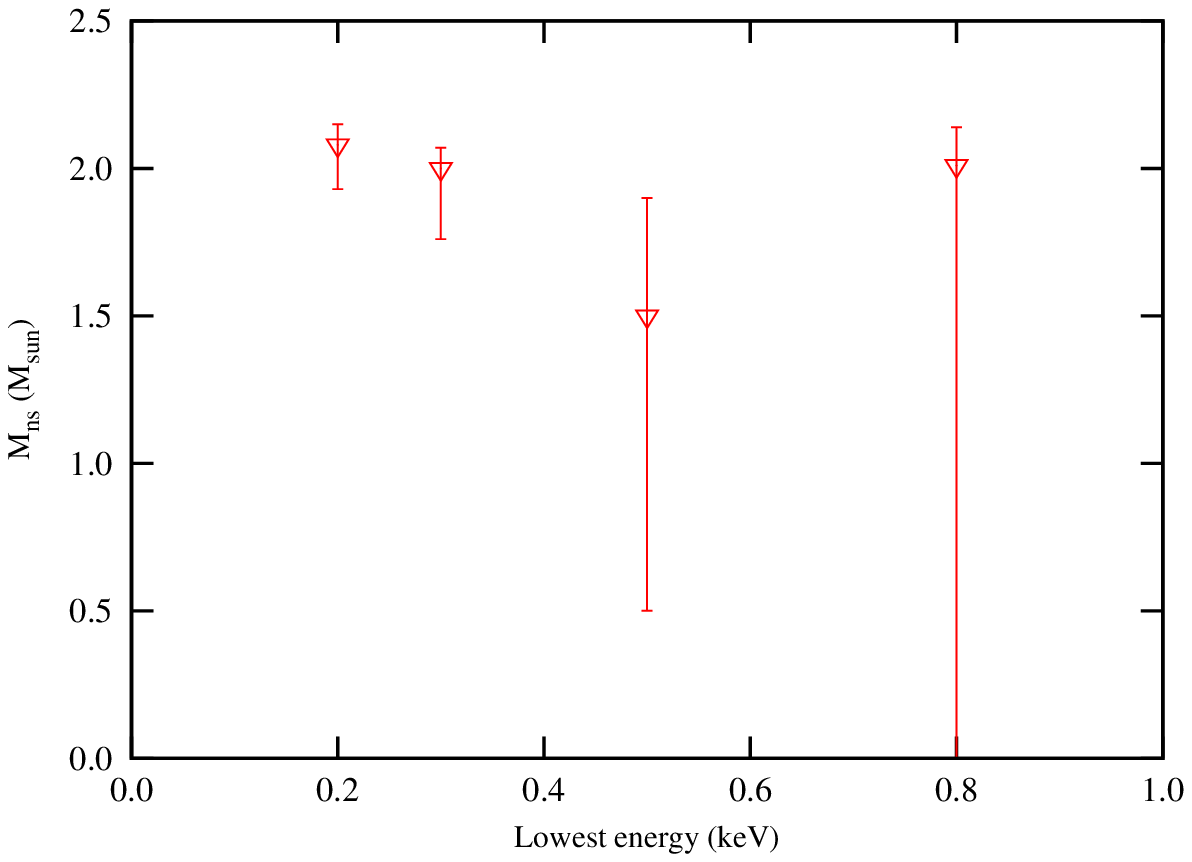}}
\subfigure{\includegraphics[width=8cm]{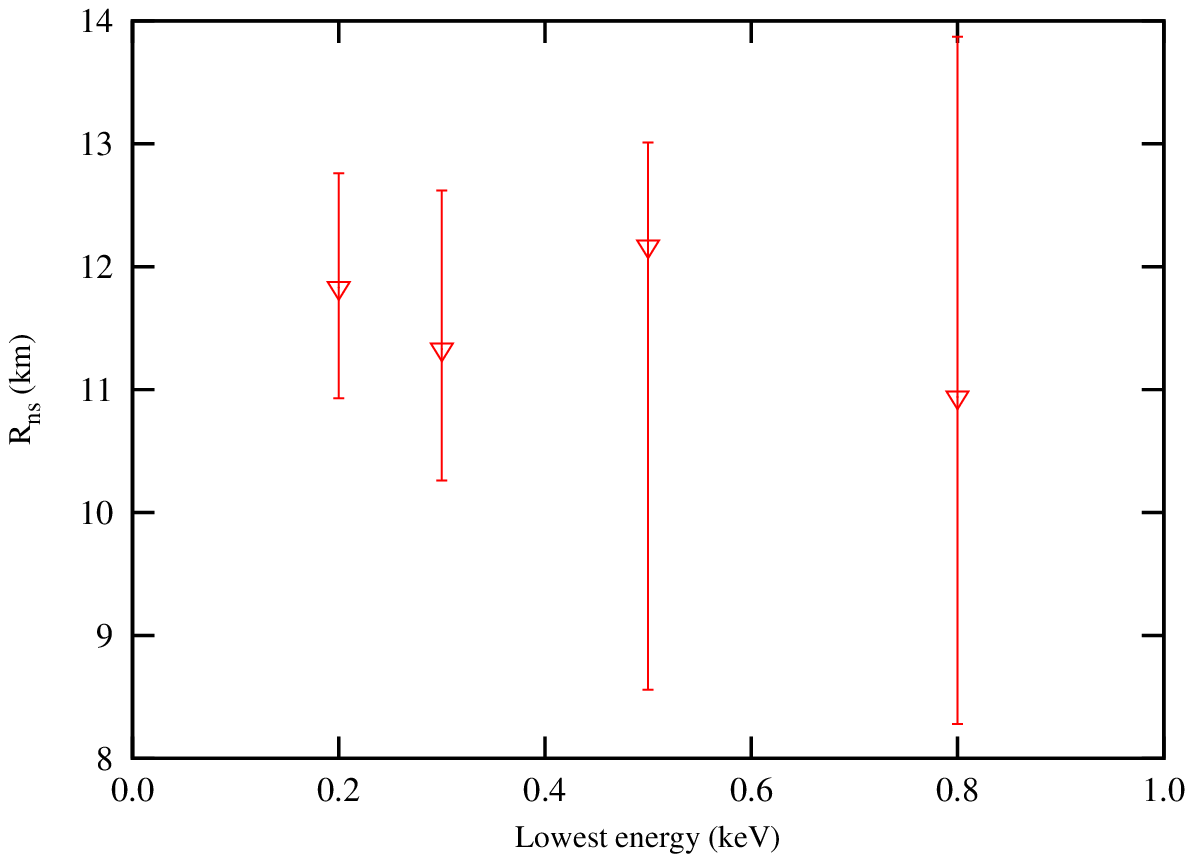}}
\caption{
The best fitting-mass (left panel) and radius (right panel) of EXO 0748$-$676 obtained from fits with the model {\sc const*phabs*nsatmos} (distance fixed to 7.1 kpc) when the minimum energy of the fits was, respectively, 0.2 keV, 0.3 keV, 0.5 keV and 0.8 keV.
The best-fitting mass values are 2.08$^{+0.07}_{-0.15}$ M$_{\odot}$, 2.00$^{+0.07}_{-0.24}$ M$_{\odot}$, 1.50$^{+0.40}_{-1.00}$ M$_{\odot}$ and 2.01$^{+0.13}_{-2.01}$ M$_{\odot}$, respectively. The best-fitting values of radius are 11.8$\pm$0.9 km, 11.3$^{+1.3}_{-1.0}$ km, 12.2$^{+0.8}_{-3.6}$ km and 10.9$^{+3.0}_{-2.6}$ km, respectively. 
}
\label{mr3}
\end{center}
\end{figure*}

\begin{figure}
\begin{center}
\includegraphics[angle=270,width=9cm]{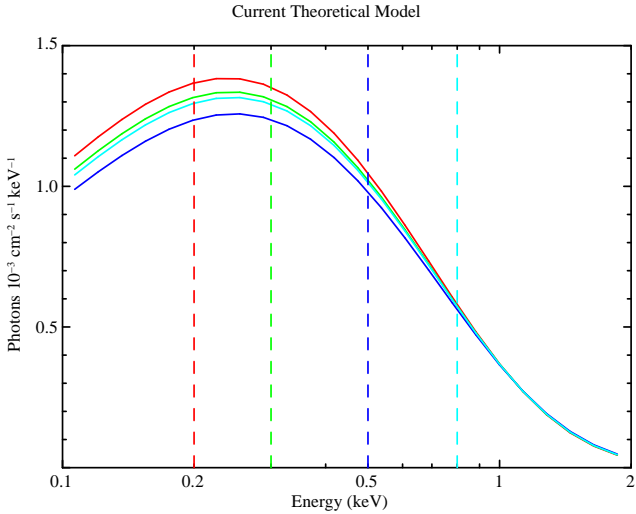}
\caption{
The {\sc nsatmos} model spectrum for the four values of the mass and radius obtained from the fits shown in Figure 5. From top to bottom the models correspond to the fits with the lowest energy used in the fits equal to 0.2 keV (red), 0.3 keV (green), 0.8 keV (cyan) and 0.5 keV (blue). The models have been normalised so that the flux above 0.5 keV is the same for all of them.
}
\label{model}
\end{center}
\end{figure}

Our best-fitting mass is consistent with the value obtained by \cite{ozel2006}, $M_{\rm ns}=2.10 \pm 0.28~M_{\odot}$, based on the measurement of the Eddington luminosity at the peak of photospheric radius expansion bursts \citep{wolff2005} and redshifted absorption lines from the NS surface during bursts \citep{cottam2002} (this spectral features could not be reproduced in a following observation, \citealt{cottam2008}), and with the dynamical constraints of \citet{Munoz2009}, $1 \le M_{\rm ns} \le 2.40~M_{\odot}$, and  \citet{Bassa2009}, $M_{\rm ns} \ge  1.27~M_{\odot}$ .

From fits to the EPIC plus RGS spectra of the first three observations presented here, plus the first XMM-Newton observation of EXO 0748$-$676 in quiescence (see \S\ref{data}), and for a distance of 7.1 kpc, \cite{diaz2011} found $M_{\rm ns}=1.77^{+0.4}_{-0.7}~M_{\odot}$ and $R_{\rm ns}=13.7 \pm 1.8$ km. On the other hand, from fits to EPIC data of the first XMM-Newton observation only, also for a distance of 7.1 kpc, \cite{guobao2011} found $M_{\rm ns}=1.55 \pm 0.12~M_{\odot}$ and $R_{\rm ns}=16.0^{+0.7}_{-1.3}$ km. We showed in \S\ref{results} that the best-fitting neutron-star mass does not change significantly whether we include the RGS data or not in the fits. The difference between our results and those of \cite{diaz2011} and \cite{guobao2011} must therefore come from the fact that the first XMM-Newton observation of EXO 0748$-$676 in quiescence requires a power-law component in the model, whereas no power-law component is needed to fit the four observations that we used here. A (mathematical) power law, as the one used in \cite{diaz2011} and \cite{guobao2011}, extends all the way down to zero, whereas the true hard emission in this source, if present, should not extend below energies comparable to the neutron-star temperature. It is then possible that the difference in the neutron-star mass between this work and those \cite{diaz2011} and \cite{guobao2011} is due to the fact that in those other papers part of the emission from the neutron-star atmosphere was actually attributed to the power-law component. 

Such a massive neutron star would in principle set very stringent constraints on the equation of state of nuclear matter (e.g., \citealt{lattimer2004}). One should then review the assumptions that lead to this result to asses its reliability. The best-fitting mass and radius depend strongly on the assumed distance to the neutron star. Following \cite{guobao2011} and \cite{diaz2011}, who used the distance from \cite{galloway2008a}, here we fixed the distance to 7.1 kpc; however, this number needs to be examined carefully. 

The distance to EXO 0748$-$676 is based on the measured peak flux of a number of PRE bursts \citep{galloway2008a} under specific assumptions (see below) that may not be valid or may contradict the best-fitting mass and radius obtained assuming that distance. More specifically, to compute the Eddington luminosity (including the relativistic corrections) of the neutron stars in X-ray bursters, and to compare those luminosities with the peak flux of the PRE bursts in these sources, \cite{galloway2008a} assumed a 1.4 $M_\odot$ and 10-km neutron-star. These values are clearly inconsistent with the best-fitting values that we give here, and would indicate that the distance should be calculated iteratively. This is, unfortunately, not straightforward, because distances determined from PRE bursts depend also upon the hydrogen mass fraction, $X$, in the burst fuel, which is normally not known. Withal, the distances given in \cite{galloway2008a}, including the distance to EXO 0748$-$676, were computed under the assumption that the emission at peak of the bursts was isotropic (cf., \citealt{lapidus1985}, \citealt{fujimoto1988}). EXO 0748$-$676 is a high-inclination system (see \S 1), and hence anisotropy will be very important in this case.

\cite{galloway2008a} found that, for a 1.4 $M_\odot$ and 10-km neutron star, and for isotropic emission at the peak of the bursts, the distance to EXO 0748$-$676 can range between $7.4 \pm 0.9$ kpc for $X=0$ and $5.7 \pm 0.7$ kpc for $X=0.7$. If we, for instance, take the lowest value in that range, the best-fitting neutron-star mass and radius in EXO 0748$-$676 are $\sim 1.6$ $M_\odot$ and $\sim 6$ km, respectively. 
Furthermore, measurements of the neutron-star bolometric flux (and hence radius) through fits of the thermal emission of neutron stars with moderately high spin frequencies, like the neutron star in EXO 0748$-$676 that has a spin frequency of 552 Hz \citep{galloway2010}, are affected by Doppler shift and frame dragging. Using eq. 27 in \cite{baubock2015}, our best-fitting neutron-star radius would be about 2 per cent larger than inferred under the assumption that the star is not spinning. This, however, is the correction averaged over all possible inclination angles of the rotation axis of the neutron star with respect to the line of sight, whereas for high-inclination systems, as is likely the case of the neutron star in EXO 0748$-$676, The correction could be as large as 12 per cent \citep{baubock2015}. All the above shows that systematic errors in the deduced neutron-star mass and radius are much larger than the statistical errors in our fits, and hence it would be disingenuous to take the best-fitting values that we obtain here at face value to draw strong conclusions about the nature of the interior of the neutron star in this system. 

To investigate the effect of the minimum energy, $E_{\rm min}$, used in the spectral fits upon the deduced mass and radius, in Figure 5 we show the best-fitting NS mass and radius of EXO 0748$-$676 using the model {\sc const*phabs*nsatmos} for, respectively, $E_{\rm min}$ equal to 0.2 keV, 0.3 keV, 0.5 keV and 0.8 keV. This Figure shows that the best-fitting NS mass decreases and the error in the mass increases as $E_{\rm min}$ increases. The best-fitting mass and error do not change significantly if $E_{\rm min}$ increases from 0.2 keV to 0.3 keV, but the error increases significantly if $E_{\rm min}$ increases from 0.3 keV to 0.5 keV. On the contrary, within errors, the best-fitting radius is independent of $E_{\rm min}$ when $E_{\rm min}$ varies between 0.2 keV and 0.8 keV, whereas the error in the best-fitting radius increases significantly as $E_{\rm min}$ increases from 0.2 keV to 0.5 keV. To understand this, in Figure 6 we plot the flux density of the {\sc nsatmos} model vs. energy for the four different values of the best-fitting mass shown in Figure 5. We normalised the models such that the flux above 0.5 keV is the same in all 4 cases, which is equivalent to what one would get by fitting these four models to data above 0.5 keV. The vertical lines in this Figure show the 4 values of $E_{\rm min}$ used in the fits to produce the plots in Figure 5. From Figure 6 it is apparent that the difference between the models increases as the energy decreases. Since the sensitivity to distinguish between two models with different values of the mass depends upon the area in between the curves of those two models, the sensitivity to the NS mass increases as $E_{\rm min}$ decreases, consistent with the fact that the errors in the NS mass decrease as $E_{\rm min}$ decreases (Figure 5). Extending the fits down to the lowest possible energy is therefore crucial to accurately measure the NS mass using the {\sc nsatmos} model, and it is specially important in the case of EXO 0748$-$676 because the column density of the interstellar material toward the source is relatively low, therefore allowing us to observe the source down to very low energies. This must be balanced, however, with the lowest energy at which the instruments are accurately calibrated. 
\cite{sartore2012} fitted the spectra of the isolated neutron star RX J1856.5$-$3754 using observations spanning a period of 10 years, and they showed that the calibration of the EPIC cameras on board XMM-Newton is accurate down to 5\% at 0.2 keV, and down to 3\% at 0.5 keV (see their Figure 1). A 5\% uncertainty in the calibration would add an extra uncertainty of $\sim 0.1~M_{\odot}$ to the neutron star mass, which is much smaller than the statistical error obtained from the fits down to 0.5 keV, therefore justifying using data down to 0.2 keV.


\section*{Acknowledgments}

This work is based on observations obtained with XMM-Newton, 
an ESA science mission with instruments and contributions directly funded by ESA Member States and the USA (NASA).
This research made use of NASA's Astrophysics Data System. 
ZC acknowledges support by the Erasmus Mundus programme.
MM wishes to thank MNM. We thanks Laurence Boirin for commenting on the manuscript. 
We are very grateful to Matteo Guainazzi for helpful comments regarding the calibration of the EPIC instruments on board XMM-Newton and the correction for background flares. 
We thank the referee, Sebastien Guillot, for his careful reading of the manuscript and the very useful suggestions that helped us improve this paper. 

This research has been funded with support from the European Commission. This publication reflects the views only of the author, and the Commission cannot be held responsible for any use which may be made of the information contained therein.

\bibliographystyle{mn2e}
\bibliography{library}

\begin{thebibliography}{60}
\expandafter\ifx\csname natexlab\endcsname\relax\def\natexlab#1{#1}\fi

\bibitem[{{Anders} \& {Grevesse}(1989)}]{anders1989}
{Anders} E., {Grevesse} N., 1989, \gca, 53, 197

\bibitem[{{Balucinska-Church} \& {McCammon}(1992)}]{balucinska1992}
{Balucinska-Church} M., {McCammon} D., 1992, \apj, 400, 699

\bibitem[{{Bassa} {et~al}\mbox{.}(2009){Bassa}, {Jonker}, {Steeghs}, \&
  {Torres}}]{Bassa2009}
{Bassa} C.~G., {Jonker} P.~G., {Steeghs} D., {Torres} M.~A.~P., 2009, \mnras,
  399, 2055

\bibitem[{{Baub{\"o}ck} {et~al}\mbox{.}(2015){Baub{\"o}ck}, {{\"O}zel},
  {Psaltis}, \& {Morsink}}]{baubock2015}
{Baub{\"o}ck} M., {{\"O}zel} F., {Psaltis} D., {Morsink} S.~M., 2015, \apj,
  799, 22

\bibitem[{{Brown} \& {Cumming}(2009)}]{brown2009}
{Brown} E.~F., {Cumming} A., 2009, \apj, 698, 1020

\bibitem[{{Cackett} {et~al}\mbox{.}(2010{\natexlab{a}}){Cackett}, {Brown},
  {Cumming}, {Degenaar}, {Miller}, \& {Wijnands}}]{cackett2010}
{Cackett} E.~M., {Brown} E.~F., {Cumming} A., {Degenaar} N., {Miller} J.~M.,
  {Wijnands} R., 2010{\natexlab{a}}, \apjl, 722, L137

\bibitem[{{Cackett} {et~al}\mbox{.}(2010{\natexlab{b}}){Cackett}, {Brown},
  {Miller}, \& {Wijnands}}]{cackett2010a}
{Cackett} E.~M., {Brown} E.~F., {Miller} J.~M., {Wijnands} R.,
  2010{\natexlab{b}}, \apj, 720, 1325

\bibitem[{{Campana} {et~al}\mbox{.}(1998){Campana}, {Colpi}, {Mereghetti},
  {Stella}, \& {Tavani}}]{campana1998}
{Campana} S., {Colpi} M., {Mereghetti} S., {Stella} L., {Tavani} M., 1998,
  \aapr, 8, 279

\bibitem[{{Cottam} {et~al}\mbox{.}(2001){Cottam}, {Kahn}, {Brinkman}, {den
  Herder}, \& {Erd}}]{cottam2001}
{Cottam} J., {Kahn} S.~M., {Brinkman} A.~C., {den Herder} J.~W., {Erd} C.,
  2001, \aap, 365, L277

\bibitem[{{Cottam}, {Paerels} \& {Mendez}(2002){Cottam}, {Paerels}, \&
  {Mendez}}]{cottam2002}
{Cottam} J., {Paerels} F., {Mendez} M., 2002, \nat, 420, 51

\bibitem[{{Cottam} {et~al}\mbox{.}(2008){Cottam}, {Paerels}, {M{\'e}ndez},
  {Boirin}, {Lewin}, {Kuulkers}, \& {Miller}}]{cottam2008}
{Cottam} J., {Paerels} F., {M{\'e}ndez} M., {Boirin} L., {Lewin} W.~H.~G.,
  {Kuulkers} E., {Miller} J.~M., 2008, \apj, 672, 504

\bibitem[{{Cox} \& {Tucker}(1969)}]{cox1969}
{Cox} D.~P., {Tucker} W.~H., 1969, \apj, 157, 1157

\bibitem[{{Degenaar} {et~al}\mbox{.}(2014){Degenaar}, {Medin}, {Cumming},
  {Wijnands}, {Wolff}, {Cackett}, {Miller}, {Jonker}, {Homan}, \&
  {Brown}}]{degenaar2014}
{Degenaar} N. {et~al.}, 2014, \apj, 791, 47

\bibitem[{{Degenaar} {et~al}\mbox{.}(2009){Degenaar}, {Wijnands}, {Wolff},
  {Ray}, {Wood}, {Homan}, {Lewin}, {Jonker}, {Cackett}, {Miller}, \&
  {Brown}}]{degenaar2009}
{Degenaar} N. {et~al.}, 2009, \mnras, 396, L26

\bibitem[{{Degenaar} {et~al}\mbox{.}(2011){Degenaar}, {Wolff}, {Ray}, {Wood},
  {Homan}, {Lewin}, {Jonker}, {Cackett}, {Miller}, {Brown}, \&
  {Wijnands}}]{degenaar2011}
{Degenaar} N. {et~al.}, 2011, \mnras, 412, 1409

\bibitem[{{den Herder} {et~al}\mbox{.}(2001){den Herder}, {Brinkman}, {Kahn},
  {Branduardi-Raymont}, {Thomsen}, {Aarts}, {Audard}, {Bixler}, {den Boggende},
  {Cottam}, {Decker}, {Dubbeldam}, {Erd}, {Goulooze}, {G{\"u}del}, {Guttridge},
  {Hailey}, {Janabi}, {Kaastra}, {de Korte}, {van Leeuwen}, {Mauche},
  {McCalden}, {Mewe}, {Naber}, {Paerels}, {Peterson}, {Rasmussen}, {Rees},
  {Sakelliou}, {Sako}, {Spodek}, {Stern}, {Tamura}, {Tandy}, {de Vries},
  {Welch}, \& {Zehnder}}]{Denherder2001}
{den Herder} J.~W. {et~al.}, 2001, \aap, 365, L7

\bibitem[{{D{\'{\i}}az Trigo} {et~al}\mbox{.}(2011){D{\'{\i}}az Trigo},
  {Boirin}, {Costantini}, {M{\'e}ndez}, \& {Parmar}}]{diaz2011}
{D{\'{\i}}az Trigo} M., {Boirin} L., {Costantini} E., {M{\'e}ndez} M., {Parmar}
  A., 2011, \aap, 528, A150

\bibitem[{{Fridriksson} {et~al}\mbox{.}(2010){Fridriksson}, {Homan},
  {Wijnands}, {M{\'e}ndez}, {Altamirano}, {Cackett}, {Brown}, {Belloni},
  {Degenaar}, \& {Lewin}}]{fridriksson2010}
{Fridriksson} J.~K. {et~al.}, 2010, \apj, 714, 270

\bibitem[{{Fujimoto}(1988)}]{fujimoto1988}
{Fujimoto} M.~Y., 1988, \apj, 324, 995

\bibitem[{{Galloway} {et~al}\mbox{.}(2010){Galloway}, {Lin}, {Chakrabarty}, \&
  {Hartman}}]{galloway2010}
{Galloway} D.~K., {Lin} J., {Chakrabarty} D., {Hartman} J.~M., 2010, \apjl,
  711, L148

\bibitem[{{Galloway} {et~al}\mbox{.}(2008){Galloway}, {Muno}, {Hartman},
  {Psaltis}, \& {Chakrabarty}}]{galloway2008a}
{Galloway} D.~K., {Muno} M.~P., {Hartman} J.~M., {Psaltis} D., {Chakrabarty}
  D., 2008, \apjs, 179, 360

\bibitem[{{Galloway}, {{\"O}zel} \& {Psaltis}(2008){Galloway}, {{\"O}zel}, \&
  {Psaltis}}]{galloway2008b}
{Galloway} D.~K., {{\"O}zel} F., {Psaltis} D., 2008, \mnras, 387, 268

\bibitem[{{Garcia} \& {Callanan}(1999)}]{garcia1999}
{Garcia} M.~R., {Callanan} P.~J., 1999, \aj, 118, 1390

\bibitem[{{Gottwald} {et~al}\mbox{.}(1986){Gottwald}, {Haberl}, {Parmar}, \&
  {White}}]{gottwald1986}
{Gottwald} M., {Haberl} F., {Parmar} A.~N., {White} N.~E., 1986, \apj, 308, 213

\bibitem[{{Haensel} \& {Zdunik}(2008)}]{haensel2008}
{Haensel} P., {Zdunik} J.~L., 2008, \aap, 480, 459

\bibitem[{{Heinke} {et~al}\mbox{.}(2006){Heinke}, {Rybicki}, {Narayan}, \&
  {Grindlay}}]{heinke2006}
{Heinke} C.~O., {Rybicki} G.~B., {Narayan} R., {Grindlay} J.~E., 2006, \apj,
  644, 1090

\bibitem[{{Homan} {et~al}\mbox{.}(2014){Homan}, {Fridriksson}, {Wijnands},
  {Cackett}, {Degenaar}, {Linares}, {Lin}, \& {Remillard}}]{homan2014}
{Homan} J., {Fridriksson} J.~K., {Wijnands} R., {Cackett} E.~M., {Degenaar} N.,
  {Linares} M., {Lin} D., {Remillard} R.~A., 2014, \apj, 795, 131

\bibitem[{{Homan} \& {van der Klis}(2000)}]{homan2000}
{Homan} J., {van der Klis} M., 2000, \apj, 539, 847

\bibitem[{{Homan}, {Wijnands} \& {van den Berg}(2003){Homan}, {Wijnands}, \&
  {van den Berg}}]{homan2003}
{Homan} J., {Wijnands} R., {van den Berg} M., 2003, \aap, 412, 799

\bibitem[{{Hynes} \& {Jones}(2009)}]{hynes2009}
{Hynes} R.~I., {Jones} E.~D., 2009, \apjl, 697, L14

\bibitem[{{Lapidus} \& {Sunyaev}(1985)}]{lapidus1985}
{Lapidus} I.~I., {Sunyaev} R.~A., 1985, \mnras, 217, 291

\bibitem[{{Lattimer} \& {Prakash}(2004)}]{lattimer2004}
{Lattimer} J.~M., {Prakash} M., 2004, Science, 304, 536

\bibitem[{{Maitra} {et~al}\mbox{.}(2011){Maitra}, {Miller}, {Raymond}, \&
  {Reynolds}}]{maitra2011}
{Maitra} D., {Miller} J.~M., {Raymond} J.~C., {Reynolds} M.~T., 2011, \apjl,
  743, L11

\bibitem[{{Medin} \& {Cumming}(2011)}]{medin2011}
{Medin} Z., {Cumming} A., 2011, \apj, 730, 97

\bibitem[{{Medin} \& {Cumming}(2014)}]{medin2014}
{Medin} Z., {Cumming} A., 2014, \apjl, 783, L3

\bibitem[{{Menou} \& {McClintock}(2001)}]{menou2001}
{Menou} K., {McClintock} J.~E., 2001, \apj, 557, 304

\bibitem[{{Mikles} \& {Hynes}(2012)}]{mikles2012}
{Mikles} V.~J., {Hynes} R.~I., 2012, \apj, 750, 132

\bibitem[{{Mu{\~n}oz-Darias} {et~al}\mbox{.}(2009){Mu{\~n}oz-Darias},
  {Casares}, {O'Brien}, {Steeghs}, {Mart{\'{\i}}nez-Pais}, {Cornelisse}, \&
  {Charles}}]{Munoz2009}
{Mu{\~n}oz-Darias} T., {Casares} J., {O'Brien} K., {Steeghs} D.,
  {Mart{\'{\i}}nez-Pais} I.~G., {Cornelisse} R., {Charles} P.~A., 2009, \mnras,
  394, L136

\bibitem[{{Narayan} \& {Yi}(1994)}]{narayan1994}
{Narayan} R., {Yi} I., 1994, \apjl, 428, L13

\bibitem[{{{\"O}zel}(2006)}]{ozel2006}
{{\"O}zel} F., 2006, \nat, 441, 1115

\bibitem[{{Parmar} {et~al}\mbox{.}(1986){Parmar}, {White}, {Giommi}, \&
  {Gottwald}}]{parmar1986}
{Parmar} A.~N., {White} N.~E., {Giommi} P., {Gottwald} M., 1986, \apj, 308, 199

\bibitem[{{Pearson} {et~al}\mbox{.}(2006){Pearson}, {Hynes}, {Steeghs},
  {Jonker}, {Haswell}, {King}, {O'Brien}, {Nelemans}, \&
  {M{\'e}ndez}}]{pearson2006}
{Pearson} K.~J. {et~al.}, 2006, \apj, 648, 1169

\bibitem[{{Piconcelli} {et~al}\mbox{.}(2004){Piconcelli}, {Jimenez-Bail{\'o}n},
  {Guainazzi}, {Schartel}, {Rodr{\'{\i}}guez-Pascual}, \&
  {Santos-Lle{\'o}}}]{piconcelli2004}
{Piconcelli} E., {Jimenez-Bail{\'o}n} E., {Guainazzi} M., {Schartel} N.,
  {Rodr{\'{\i}}guez-Pascual} P.~M., {Santos-Lle{\'o}} M., 2004, \mnras, 351,
  161

\bibitem[{{Protassov} {et~al}\mbox{.}(2002){Protassov}, {van Dyk}, {Connors},
  {Kashyap}, \& {Siemiginowska}}]{protassov2002}
{Protassov} R., {van Dyk} D.~A., {Connors} A., {Kashyap} V.~L., {Siemiginowska}
  A., 2002, \apj, 571, 545

\bibitem[{{Ratti} {et~al}\mbox{.}(2012){Ratti}, {Steeghs}, {Jonker}, {Torres},
  {Bassa}, \& {Verbunt}}]{Ratti2012}
{Ratti} E.~M., {Steeghs} D.~T.~H., {Jonker} P.~G., {Torres} M.~A.~P., {Bassa}
  C.~G., {Verbunt} F., 2012, \mnras, 420, 75

\bibitem[{{Sartore} {et~al}\mbox{.}(2012){Sartore}, {Tiengo}, {Mereghetti}, {De
  Luca}, {Turolla}, \& {Haberl}}]{sartore2012}
{Sartore} N., {Tiengo} A., {Mereghetti} S., {De Luca} A., {Turolla} R.,
  {Haberl} F., 2012, \aap, 541, A66

\bibitem[{{Schatz} {et~al}\mbox{.}(2014){Schatz}, {Gupta}, {M{\"o}ller},
  {Beard}, {Brown}, {Deibel}, {Gasques}, {Hix}, {Keek}, {Lau}, {Steiner}, \&
  {Wiescher}}]{schatz2014}
{Schatz} H. {et~al.}, 2014, \nat, 505, 62

\bibitem[{{Sidoli}, {Parmar} \& {Oosterbroek}(2005){Sidoli}, {Parmar}, \&
  {Oosterbroek}}]{sidoli2005}
{Sidoli} L., {Parmar} A.~N., {Oosterbroek} T., 2005, \aap, 429, 291

\bibitem[{{Smith} {et~al}\mbox{.}(2001){Smith}, {Brickhouse}, {Liedahl}, \&
  {Raymond}}]{smith2001}
{Smith} R.~K., {Brickhouse} N.~S., {Liedahl} D.~A., {Raymond} J.~C., 2001,
  \apjl, 556, L91

\bibitem[{{Str{\"u}der} {et~al}\mbox{.}(2001){Str{\"u}der}, {Briel}, {Dennerl},
  {Hartmann}, {Kendziorra}, {Meidinger}, {Pfeffermann}, {Reppin}, {Aschenbach},
  {Bornemann}, {Br{\"a}uninger}, {Burkert}, {Elender}, {Freyberg}, {Haberl},
  {Hartner}, {Heuschmann}, {Hippmann}, {Kastelic}, {Kemmer}, {Kettenring},
  {Kink}, {Krause}, {M{\"u}ller}, {Oppitz}, {Pietsch}, {Popp}, {Predehl},
  {Read}, {Stephan}, {St{\"o}tter}, {Tr{\"u}mper}, {Holl}, {Kemmer}, {Soltau},
  {St{\"o}tter}, {Weber}, {Weichert}, {von Zanthier}, {Carathanassis}, {Lutz},
  {Richter}, {Solc}, {B{\"o}ttcher}, {Kuster}, {Staubert}, {Abbey}, {Holland},
  {Turner}, {Balasini}, {Bignami}, {La Palombara}, {Villa}, {Buttler},
  {Gianini}, {Lain{\'e}}, {Lumb}, \& {Dhez}}]{struder2001}
{Str{\"u}der} L. {et~al.}, 2001, \aap, 365, L18

\bibitem[{{van Peet} {et~al}\mbox{.}(2009){van Peet}, {Costantini},
  {M{\'e}ndez}, {Paerels}, \& {Cottam}}]{vanpeet2009}
{van Peet} J.~C.~A., {Costantini} E., {M{\'e}ndez} M., {Paerels} F.~B.~S.,
  {Cottam} J., 2009, \aap, 497, 805

\bibitem[{{Verner} {et~al}\mbox{.}(1996){Verner}, {Ferland}, {Korista}, \&
  {Yakovlev}}]{verner1996}
{Verner} D.~A., {Ferland} G.~J., {Korista} K.~T., {Yakovlev} D.~G., 1996, \apj,
  465, 487

\bibitem[{{Villarreal} \& {Strohmayer}(2004)}]{villarreal2004}
{Villarreal} A.~R., {Strohmayer} T.~E., 2004, \apjl, 614, L121

\bibitem[{{Wilms}, {Allen} \& {McCray}(2000){Wilms}, {Allen}, \&
  {McCray}}]{wilms2000}
{Wilms} J., {Allen} A., {McCray} R., 2000, \apj, 542, 914

\bibitem[{{Wolff} {et~al}\mbox{.}(2008){Wolff}, {Ray}, {Wood}, \&
  {Wijnands}}]{wolff2008b}
{Wolff} M., {Ray} P., {Wood} K., {Wijnands} R., 2008, The Astronomer's
  Telegram, 1812, 1

\bibitem[{{Wolff} {et~al}\mbox{.}(2005){Wolff}, {Becker}, {Ray}, \&
  {Wood}}]{wolff2005}
{Wolff} M.~T., {Becker} P.~A., {Ray} P.~S., {Wood} K.~S., 2005, \apj, 632, 1099

\bibitem[{{Wolff}, {Ray} \& {Wood}(2008){Wolff}, {Ray}, \& {Wood}}]{wolff2008a}
{Wolff} M.~T., {Ray} P.~S., {Wood} K.~S., 2008, The Astronomer's Telegram,
  1736, 1

\bibitem[{{Yakovlev} \& {Pethick}(2004)}]{yakovlev2004}
{Yakovlev} D.~G., {Pethick} C.~J., 2004, \araa, 42, 169

\bibitem[{{Yan}, {Sadeghpour} \& {Dalgarno}(1998){Yan}, {Sadeghpour}, \&
  {Dalgarno}}]{yan1998}
{Yan} M., {Sadeghpour} H.~R., {Dalgarno} A., 1998, \apj, 496, 1044

\bibitem[{{Zhang} {et~al}\mbox{.}(2011){Zhang}, {M{\'e}ndez}, {Jonker}, \&
  {Hiemstra}}]{guobao2011}
{Zhang} G., {M{\'e}ndez} M., {Jonker} P., {Hiemstra} B., 2011, \mnras, 414,
  1077

\end{thebibliography}



\vspace{1cm}
\footnotesize{This paper was typeset using a \LaTeX\ file prepared by the 
author}


\end{document}